\documentclass[useAMS,usenatbib]{mn2e}

\usepackage[T1]{fontenc}
\usepackage{aecompl}
\usepackage{graphicx}
\usepackage{array}
\usepackage{lscape} 
\usepackage{txfonts}
\usepackage{longtable}
\usepackage{color}
\usepackage{ulem}    
\usepackage{times}
\usepackage{subfigure}
\usepackage{booktabs}
\newcommand{\dens}{$n_{\rm e}$}
\newcommand{\temp}{$T_{\rm e}$}

\newcommand\ion[2]{#1$\;${\scshape{#2}}}
\newcommand{\Msun}{M$_\odot$}

\newcommand{\cmtres}{$\rm{cm}^{-3}$}

\newcommand{\abhei}{$\mbox{He}^{+}$}
\newcommand{\abheii}{$\mbox{He}^{++}$}
\newcommand{\aboii}{$\mbox{O}^{+}$}
\newcommand{\aboiii}{$\mbox{O}^{++}$}
\newcommand{\abnii}{$\mbox{N}^{+}$}

\newcommand{\abneiii}{$\mbox{Ne}^{++}$}

\newcommand{\absii}{$\mbox{S}^{+}$}
\newcommand{\absiii}{$\mbox{S}^{++}$}

\newcommand{\abariii}{$\mbox{Ar}^{++}$}
\newcommand{\abariv}{$\mbox{Ar}^{+3}$}

\newcommand{\abcii}{$\mbox{C}^{++}$}

\newcommand{\abcliii}{$\mbox{Cl}^{++}$}

\title[O enrichment in C-rich PNe]{Oxygen enrichment in carbon-rich planetary nebulae}
\author[Delgado-Inglada et al.]{Gloria Delgado-Inglada$^{1}$\thanks{E-mail: gloria.delgado.inglada@gmail.com (GDI)}, 
M\'onica Rodr\'iguez$^{2}$, Manuel Peimbert$^{1}$, Gra\.zyna Stasi\'nska$^{3}$ 
\newauthor and Christophe Morisset$^{1}$\\
$^{1}$Instituto de Astronom\'ia, Universidad Nacional Aut\'onoma de M\'exico, Apdo. Postal 70264,04510, M\'exico D. F., Mexico\\
$^{2}$Instituto Nacional de Astrof\'isica, \'Optica y Electr\'onica (INAOE), Apdo. Postal 51 y 216, 72000 Puebla, Pue. Mexico\\
$^{3}$LUTH, Observatoire de Paris, CNRS, Universit\'e Paris Diderot; Place Jules Janssen, F-92190 Meudon, France}

\begin{document}

\date{}

\pagerange{\pageref{firstpage}--\pageref{lastpage}} \pubyear{2014}

\maketitle

\label{firstpage}

\begin{abstract}
We study the relation between the chemical composition and the type of dust present in a group
of 20 Galactic planetary nebulae (PNe) that have high quality optical and infrared spectra. The 
optical spectra are used, together with the best available ionization correction factors, to
calculate  the abundances of Ar, C, Cl, He, N, Ne, and O relative to H. The infrared spectra are used to
classify the PNe in two groups depending on whether the observed dust features are
representative of oxygen-rich or carbon-rich environments. The sample contains one object from
the halo, eight from the bulge, and eleven from the local disc. We compare their chemical
abundances with nucleosynthesis model predictions and with the ones obtained in seven Galactic
\ion{H}{ii} regions of the solar neighbourhood. 

We find evidence of O enrichment (by $\sim0.3$ dex) in all but one of the PNe with carbon-rich 
dust (CRD). Our analysis shows that Ar, and especially Cl, are the best metallicity indicators
of the progenitors of PNe. There is a tight correlation between the abundances of Ar and Cl in
all the objects, in agreement with a lockstep evolution of both elements. The range of
metallicities implied by the Cl abundances covers one order of magnitude and we find
significant  differences in the initial masses and metallicities of the PNe with CRD and
oxygen-rich dust (ORD). The PNe with CRD tend to have intermediate masses and low metallicities,
whereas most of the PNe with ORD show higher enrichments in N and He, suggesting that they had
high-mass progenitors.
\end{abstract}


\section{Introduction}
Planetary nebulae (PNe) are the final products of many stars with masses below $\sim 8$ $M_\odot$. 
This evolutionary stage occurs after the asymptotic giant branch (AGB) phase, if the star reaches a
temperature high enough to ionize the circumstellar gas (the AGB external envelope ejected through a
stellar wind) before it dilutes into the interstellar medium. The ionized gas contains valuable
information on the nucleosynthesis processes occurring inside the AGB stars (because some elements are
produced in the interior of these stars and then carried to the surface) and on the chemical
composition of the environment where the star was born (since other elements remain unchanged during
the life of the star). The chemical abundances in PNe can be compared with the ones computed using the
same techniques in \ion{H}{ii} regions in order to obtain information not only about the efficiency of
stellar nucleosynthesis and dredge-up processes but also about the development of chemical
evolution in different galaxies.

The oxygen abundance has been traditionally used as a proxy for the metallicity in ionized nebulae
because oxygen is the element for which more reliable abundances can be obtained. Bright emission
lines from two of its ionization states, \aboii\ and  \aboiii, can be easily measured in optical
spectra, and the correction for the contribution of higher ionization states is large only for PNe  of
very high excitation. However, AGB stars can modify the oxygen abundance via  two mechanisms. The
third dredge-up (TDU) is a mixing event that transports to the stellar surface material created by He
burning and the {\it s}-process. The material is heavily enriched in carbon but also has some oxygen
\citep[see e.g.][]{kar14}. On the other hand, hot bottom burning (HBB) occurs in the most massive
progenitors and consists in the penetration of the bottom of the convective envelope into a region
where the temperature is sufficient to activate the CNO-cycle, which can destroy oxygen \citep[see
e.g.][]{kar14}. The products are then conveyed to the surface. 

The efficiencies of the TDU and HBB depend on the initial mass of the star and on its metallicity, 
and the predictions differ from model to model. For example, standard nucleosynthesis models with no
extra mixing processes, such as those of \citet{kar10}, do not predict a significant production of
oxygen at solar or half solar  metallicity. However, models that include diffusive convective
overshooting, like those computed by  \citet{mar01} and \citet{pig13}, predict a significant
production of oxygen, even at solar metallicities. This mixing mechanism was first introduced by
\citet{her97} and causes the convective envelope to penetrate more deeply into the star and, in
combination with an efficient third dredge up, leads to an increase of the oxygen abundance for stars
between 1.5 \Msun\ and 3 \Msun. There are other mechanisms that can produce extra mixing, such as
rotation, magnetic fields, and thermohaline mixing \citep[see e.g.][]{kar14}. These mechanisms are
not well understood, but reliable determinations of the chemical composition of ionized nebulae might
help us to constrain their efficiencies. Observationally, claims of oxygen production in low mass
stars have been restricted to low metallicity PNe \citep[see e.g.][]{peq00, lei06}. \citet{rod11}
found an overabundance of oxygen in nearby PNe of near-solar metallicity when they compared them with
local \ion{H}{ii} regions, but a detailed analysis of the clues provided by other elements is required
in order to ascertain the cause of this overabundance. On the other hand, the AGB models of
\citet{kar10} and \citet{pig13} do not predict a significant destruction of oxygen for metallicities
above $Z\simeq0.01$, but the models of \citet{kar10} with $Z\simeq0.008$ show that O/H can decrease by
$\sim0.1$--0.2 dex for stars with masses $M\ge5$ \Msun, and this effect could also be more important
at lower metallicities \citep[see e.g.][]{sta10}.

Other elements whose abundances we can derive and use as proxies for nebular metallicities are sulfur,
neon, chlorine and argon. However, these elements require corrections for the contribution of
unobserved ions to their total abundances, the so-called ionization correction factors (ICFs). Up
until recently, the available ICFs were based on the similarities of ionization potentials of the ions
involved or on the results provided by a small number of photoionization models
\citep[e.g.][]{kin94}. This problem is now alleviated by the new and more reliable ICFs calculated by
\citet*{del14b}, which are based on large grids of models. These new ICFs can be used to derive more
reliable estimates of the chemical abundances, including those elements that are expected to be
modified during the evolution of the stellar progenitors of PNe, namely, helium, nitrogen, and carbon.

A very useful abundance ratio in the study of PNe is C/O. Its value is related to the initial mass
and metallicity of the progenitor star since these parameters determine the efficiencies of the
nucleosynthesis processes that change the relative abundances of C and O. Stars are born with
C/O~$<1$ (C/O~$\simeq0.5$ for solar metallicity; \citealt{all02}). The lowest mass 
AGB stars (with initial masses below $\sim1.5$ \Msun), do not go through the TDU or HBB, and 
thus keep their C/O values below 1. They are thus defined as O-rich stars. Intermediate mass stars, 
with masses between $\sim1.5$--2.0 \Msun\ and $\sim4$--5 \Msun\ (the mass range depends on metallicity
and on the assumptions of the model calculations), suffer the TDU that increases the C abundance in 
the surface transforming the O-rich star into a C-rich star; these stars have C/O~$>1$. In the 
most massive progenitors, with masses above $\sim4-5$ \Msun, HBB counteracts the effect of 
the TDU and prevents the formation of a C-rich star; therefore, for these stars one gets again
C/O~$<1$. 

Moreover, the value of C/O in the atmospheres of AGB stars, which are among the most efficient 
sources of dust in the Galaxy \citep{whi10}, defines the type of dust grains that are formed.  When
oxygen is more abundant than carbon, oxygen-rich grains, such as silicates and oxides, will  be
formed. If carbon is more abundant, other dust species are expected to form, such as  SiC, TiC, and
MgS. If we assume that the C/O abundance ratio (where C and O are the total abundances of 
these elements present in dust grains and in the gas) has not changed from the time when dust 
formation initiated, the type of grains found in PNe will be related to the value of C/O in the atmosphere 
of the PN progenitor. Therefore, using infrared spectra we can identify different dust grains in PNe 
and thus classify them as C-rich (C/O~$>1$) or O-rich (C/O~$<1$) PNe. 
This inference on whether the ejected atmosphere of the progenitor star
has C/O above or below 1 can be better than the one obtained with the C/O abundance ratio derived
using emission lines, since the latter can be affected by depletion of C or O into dust grains and by
the uncertainties involved in its calculation \citep[see e.g.][]{del14a}. 

Our aim here is to study the chemical abundances of a group of PNe that have high quality optical and 
infrared spectra. The optical spectra are used to derive the element abundances and the infrared 
spectra are used to identify PNe with either oxygen-rich dust (ORD) or carbon-rich dust (CRD).  We
explore if the two groups of PNe, characterized by different dust grains, show differences in their
chemical content. The abundances of the PNe are also compared with the predictions of recent
nucleosynthesis models and with the chemical abundances derived for a group of Galactic \ion{H}{ii}
regions.

\section[]{The sample}

For the purpose of the present study, we need a sample of PNe with high quality spectra (i.e., with a
high signal-to-noise ratio and spectral resolution better than 4 \AA), allowing us to perform the
usual plasma diagnostics with good precision, to derive accurate abundances of elements that show only
weak lines, such as Cl, and to determine (in most cases) the carbon-to-oxygen abundance ratio from
very weak recombination lines. Most of the 57 planetary nebulae compiled by \citet{del14a} meet these
requirements. From this sample, \citet{del14a} extracted 33 PNe with available infrared spectra. In
this paper, we have selected the 20 PNe that we can classify as oxygen-rich or carbon-rich, i.e.,
objects that have in their infrared spectra either ORD features (amorphous or crystalline
silicates) or the broad features around 11 and 30 $\mu$m associated with SiC and MgS, respectively,
that are expected to arise in carbon-rich environments \citep{whi03}. Many of the objects in the
sample also show features from polycyclic aromatic hydrocarbons (PAHs), but since PAHs are sometimes
detected in PNe with silicates, we do not use these features to classify the PNe as carbon-rich or
oxygen-rich. Our selection criterion excludes NGC~2392, that contains dust grains (as can be inferred
from the shape of its continuum), but shows no clear evidence of any of the previously mentioned dust
features. We also excluded from the analysis PNe like NGC~6884 and NGC~6741 that show PAH emission,
but whose available infrared spectra do not allow us to detect or to rule out the presence of
silicates or the 11 and 30 $\mu$m features.

In summary, we use the above information to classify the sample in two groups: 1) ORD PNe are those
with silicates and 2) CRD PNe are those with the broad features associated with SiC and MgS. Some of
the PNe with ORD and CRD also show emission from PAHs. The so-called dual chemistry PNe are those
having PAH emission and silicate dust features \citep[see e.g.][]{wat98}. The compilation of dust
features is taken from \citet{del14a} and is based on {\it Spitzer} and {\it ISO} spectra. 

The sample of PNe includes objects from the Galactic disc, bulge, and halo. DdDm~1, with a height
above the Galactic plane of $\sim7$ kpc \citep{qui07}, is our only object from the halo. Eight PNe
belong to the bulge: Cn~1-5, H~1-50, M~1-42, M~2-27, M~2-31, M~2-42, NGC~6439, and M~1-20.  These PNe
satisfy the three criteria used by \citet{sta91} to identify bulge PNe, namely, they are located
within 10$^\circ$ in terms of latitude and longitude with respect to the Galactic centre, their
angular diameters are smaller than 20$^{\prime\prime}$, and their radio fluxes at 6 cm are lower than
100 mJy. The remaining 11 PNe belong to the disc, and a calculation using the distances provided
by \citet{zha95} and adopting a solar Galactocentric distance of 8.0 kpc indicates that they have
Galactocentric distances between 6 and 9 kpc, so that they can be loosely described as belonging to
the solar neighbourhood. 

Table~\ref{tab:1} lists the 20 PNe studied here with the dust features identified in each of them, 
some comments about their central stars, and the references from where the optical line fluxes 
and the characteristics of the central stars have been taken. 

\renewcommand{\arraystretch}{1.2}
\begin{table}
\caption{Sample of planetary nebulae}             
\label{tab:1}      
\centering          
\begin{tabular}{lccccc}
\hline\hline   
\multicolumn{1}{c}{Object} & \multicolumn{1}{c}{Aliph. C} & \multicolumn{1}{c}{Arom. C} 
& \multicolumn{1}{c}{Silic.} & \multicolumn{1}{c}{Comments} & \multicolumn{1}{c}{Ref.} \\
\hline 
\multicolumn{6}{l}{O-rich dust PNe}\\    
\hline                
Cn~1-5       & No & Yes & Yes &  [WR] & (1, 7)\\
DdDm~1     & No & No & Yes  &  \ldots &(2)\\
H~1-50       & No & Yes & Yes &  \ldots & (1) \\
M~1-42       & No & Yes & Yes & \ldots & (3) \\
M~2-27       & No & Yes & Yes &  wels & (1, 8)\\
M~2-31       & No & Yes & Yes &  [WR] & (1, 7)\\
M~2-42       & No & No & Yes  &  wels & (1, 9)\\
MyCn~18    & No & Yes & Yes &  \ldots & (4)\\
NGC~3132 & No & No & Yes  &  binary & (4, 10)\\
NGC~6210 & No & No & Yes  &  \ldots & (5)\\
NGC~6439 & No & Yes & Yes &  \ldots & (1)\\
NGC~6543 & No & No & Yes  &  wels & (5, 7) \\
NGC~7026 & No & Yes & Yes &  [WR] & (2, 7)\\
[0.2ex] \hline \\
[-2.8ex]
\multicolumn{6}{l}{C-rich dust PNe}\\
\hline  
Hu~2-1     & Yes & Yes & No &  \ldots & (2)\\
IC~418       & Yes & Yes & No &  \ldots & (6)\\
M~1-20       & Yes & Yes & No &  wels & (1, 11)\\
NGC~40     & Yes & Yes & No &  [WR] & (5, 7) \\
NGC~3242 & Yes & No & No  &  \ldots & (4)\\
NGC~3918 & Yes & No & No  &  \ldots & (4)\\
NGC~6826 & Yes & No & No  &  binary & (5, 12)\\
[0.2ex] \hline \\
[-2.8ex]
\multicolumn{6}{l}{
\begin{minipage}{8.0cm}
\small References: (1) \citet{wan07},  (2) \citet*{wes05}, (3) \citet{liu01}, (4) \citet{tsa03}, 
(5) \citet{liu04a}, (6) \citet{sha03}, (7) \citet{tyl93}, (8) \citet{gor04}, (9) \citet{pew11},
(10) \citet{cia99}, (11) \citet{gor09}, (12) \citet{men89}.\\
\small Comments-- [WR]: Wolf Rayet central star, wels: weak emission line star.
\end{minipage}
}\\ 
\end{tabular}
\end{table}
\renewcommand{\arraystretch}{1.}

In addition to these PNe, we selected from the literature the seven Galactic \ion{H}{ii} regions from 
the solar neighbourhood with the best available optical spectra: M8, M16, M17, M20, M42, 
NGC~3576, and NGC~3603 \citep{est04,gar04,gar06,gar07}. They cover a range 
of Galactocentric distances that goes from 6.3 kpc (M16) to 8.65 kpc (NGC~3603), similar 
to the range covered by the disc PNe.

\section{The analysis}
\subsection{Physical conditions}

We use the software PyNeb \citep{lur14} to calculate the  physical conditions, the ionic abundances
from collisionally excited lines (CELs), and  the ionic abundances of \abhei\ and \abheii\ from
recombination lines (RLs). We have selected for our calculations the atomic data listed in
Table~\ref{tab:2}. 

\begin{table*}
\caption{Atomic Data}             
\label{tab:2}      
\centering          
\begin{tabular}{l l l }
\hline\hline       
Ion & Transition Probabilities & Collisional Strengths\\ 
\hline                    
\aboii   & \citet{zei82} & \citet{kis09}\\
\aboiii  & \citet{sto00} & \citet{sto14}\\
\abnii   & \citet*{gal97} & \citet{tay11}\\
\absii   & \citet{men82} & \citet{tay10}\\
\absiii  & \citet{pod09} & \citet{tay99}\\
\abneiii & \citet{gal97} & \citet{Mc00}\\
\abariii & \citet{mun09} & \citet{mun09}\\
\abariv & \citet{men82} & \citet{ram97}\\
\abcliii & \citet{men82} & \citet{kru70}\\
\hline                  
\end{tabular}
\end{table*}

We determine two electron temperatures, \temp, for each nebula using the diagnostic 
ratios [\ion{O}{iii}] $\lambda$4363/($\lambda$4959+$\lambda$5007) and 
[\ion{N}{ii}] $\lambda$5755/($\lambda$6548+$\lambda$6583). We also compute 
an average density from the available ratios among the following: 
[\ion{O}{ii}] $\lambda$3726/$\lambda$3729, [\ion{S}{ii}] $\lambda$6716/$\lambda$6731, 
[\ion{Cl}{iii}] $\lambda$5518/$\lambda$5538, and [\ion{Ar}{iv}] $\lambda$4711/$\lambda$4740. 
For each nebula, the adopted density is the median of the distribution of average values 
obtained in the Monte Carlo calculations that we use to estimate the uncertainties 
(see Section~\ref{uncert}). We prefer 
to use the median since, in the low and high density limits, it is less affected than the mean 
by the large changes in the density caused by small variations in the intensity ratios.

When N$^{++}$ is an important ionization state in a nebula, the recombination of this ion may 
contribute significantly to the intensity of the [\ion{N}{ii}] $\lambda$5755 line \citep{rub86}. If
this effect is not taken into account, the value of \temp([\ion{N}{ii}]) can be overestimated.
\citet{liu00} derived an expression to calculate this contribution that depends on the N$^{++}$ 
abundance, and also, to a small degree, on the value of \temp. The N$^{++}$ abundance can be
determined using optical RLs, ultraviolet CELs, or infrared CELs, with the choice of lines leading
to very different results \citep[see e.g.][]{liu00}. Therefore, the size of the correction is
somewhat uncertain, and we decided not to use it here. However, we estimate an upper limit to this
effect by using the largest possible value for the N$^{++}$ abundance, the one implied by RLs. The
most affected objects are the PNe DdDm~1, NGC 3242, and NGC~6826. In the case of DdDm~1, where
\citet{wes05} estimate that recombination is the dominant contribution to the emission of
[\ion{N}{ii}] $\lambda$5755, we could use \temp([\ion{O}{iii}]) to derive all the chemical abundances,
but we find very similar results. For NGC 3242 and NGC~6826, the corrected values of
\temp([\ion{N}{ii}]) increase the O/H and Cl/O abundance ratios by $\sim0.2$--0.6 dex. For the other
objects, the corrections are much smaller, leading to changes in our abundance ratios lower than
0.1 dex. These changes, even the larger ones, do not affect in a significant way the conclusions we
get from our results below and we will not consider them in what follows.

Table~\ref{tab:3} presents the derived temperatures and densities that will be used to compute 
the ionic abundances. 

\renewcommand{\arraystretch}{1.2}
\begin{table*}
\caption{Physical conditions}             
\label{tab:3}      
\centering          
\begin{tabular}{lr@{}lr@{}lr@{}lr@{}lr@{}lr@{}lr@{}l}
\hline\hline   
\multicolumn{1}{c}{Object} & \multicolumn{2}{c}{\dens([\ion{O}{ii}])} & \multicolumn{2}{c}{\dens([\ion{S}{ii}])} & \multicolumn{2}{c}{\dens([\ion{Cl}{iii}])} & \multicolumn{2}{c}{\dens([\ion{Ar}{iv}])} & \multicolumn{2}{c}{\dens(adopted)} & \multicolumn{2}{c}{\temp([\ion{N}{ii}])} & \multicolumn{2}{c}{\temp([\ion{O}{iii}])}\\
\multicolumn{1}{c}{} & \multicolumn{2}{c}{(\cmtres)} & \multicolumn{2}{c}{(\cmtres)} & \multicolumn{2}{c}{(\cmtres)} & \multicolumn{2}{c}{(\cmtres)} & 
\multicolumn{2}{c}{(\cmtres)} & \multicolumn{2}{c}{(K)} & \multicolumn{2}{c}{(K)}\\ 
\hline 
\multicolumn{7}{l}{O-rich dust PNe}\\    
\hline           
Cn~1-5    & \ldots & \ldots & 4200 & $_{-1100}^{+1700}$  & 3000 & $_{-1500}^{+2200}$  & 2280 & $_{-1500}^{+2300}$  & 3600 & $_{-1200}^{+1400}$  & 7400 & $_{-200}^{+180}$  & 8680 & $_{-140}^{+160}$  \\
DdDm~1    & 4800 & $_{-1200}^{+1600}$  & \ldots & \ldots  & \ldots & \ldots & 3900 & $_{-2400}^{+5100}$ & 4600 & $_{-1500}^{+2400}$  & 12720 & $_{-640}^{+470}$  & 12070 & $\pm320$   \\
H~1-50    & 6700 & $_{-1800}^{+2500}$ & 7500 & $_{-2400}^{+3700}$ & 8900 & $_{-3100}^{+5600}$ & 12200 & $_{-1500}^{+2300}$ & 9300 & $_{-1500}^{+2500}$ & 11280 & $_{-430}^{+460}$ & 10870 & $_{-230}^{+240}$ \\
M~1-42    & 1360 & $_{-200}^{+260}$  & 1250 & $_{-300}^{+270}$  & 1520 & $_{-400}^{+410}$  & 570 & $_{-310}^{+520}$  & 1240 & $\pm220$  & 8860 & $_{-210}^{+240}$  & 9140 & $\pm160$   \\
M~2-27    & \ldots & \ldots & 7000 & $_{-2100}^{+3500}$  & 12600 & $_{-4900}^{+9600}$  & 10700 & $_{-1400}^{+1900}$   & 10700 & $_{-2400}^{+3500}$  & 9240 & $_{-450}^{+270}$  & 8180 & $_{-130}^{+150}$  \\
M~2-31    & \ldots & \ldots  & 6100 & $_{-1500}^{+2500}$  & 6900 & $_{-2500}^{+4100}$  & 4190 & $_{-710}^{+900}$   & 5900 & $_{-1000}^{+1700}$  & 11380 & $_{-400}^{+440}$  & 9760 & $_{-180}^{+160}$   \\
M~2-42    & \ldots & \ldots  & 3420 & $_{-850}^{+1200}$  & 3200 & $_{-1700}^{+2600}$  & 4100 & $_{-2800}^{+4000}$  & 3700 & $_{-1200}^{+2200}$  & 10060 & $_{-320}^{+420}$  & 8410 & $_{-120}^{+130}$   \\
MyCn~18   & \ldots & \ldots & 5100 & $_{-1300}^{+2600}$  & 9200 & $_{-1200}^{+1900}$ & \ldots & \ldots  & 7400 & $_{-1100}^{+1300}$  & 10020 & $_{-290}^{+300}$  & 7310 & $_{-120}^{+100}$  \\
NGC~3132  & \ldots & \ldots  & 590 & $_{-180}^{+170}$  & 740 & $_{-280}^{+350}$  & 540 & $_{-360}^{+420}$ & 621 & $_{-63}^{+200}$  & 9350 & $_{-220}^{+280}$  & 9450 & $_{-180}^{+200}$ \\
NGC~6210  & \ldots & \ldots  & 4070 & $_{-950}^{+1400}$  & 3240 & $_{-940}^{+1200}$  & \ldots & \ldots & 3810 & $_{-860}^{+870}$  & 11750 & $_{-400}^{+390}$  & 9970 & $\pm290$  \\
NGC~6439  & 3740 & $_{-730}^{+1100}$  & 4700 & $_{-1100}^{+1900}$  & 5060 & $_{-900}^{+840}$  & 6070 & $_{-990}^{+1300}$ & 4980 & $_{-480}^{+670}$  & 9630 & $_{-240}^{+300}$  & 10260 & $_{-160}^{+260}$  \\
NGC~6543  & 4500 & $_{-1000}^{+1800}$  & 5900 & $_{-2100}^{+5200}$ & 4400 & $_{-1900}^{+3300}$ & 3400 & $_{-2300}^{+3200}$ & 5100 & $_{-1300}^{+2100}$  & 9930 & $_{-490}^{+490}$  & 7780 & $_{-200}^{+220}$ \\
NGC~7026  & 2910 & $_{-35}^{+32}$ & 26500 & $_{-8800}^{+13000}$  & 8800 & $_{-1300}^{+1500}$  & 5210 & $_{-710}^{+1100}$ & 6020 & $_{-780}^{+4500}$  & 9730 & $_{-450}^{+300}$  & 9130 & $_{-150}^{+130}$ \\
[0.2ex] \hline \\
[-2.8ex]
\multicolumn{7}{l}{C-rich dust PNe}\\
\hline  
Hu~2-1    & 9000 & $_{-2600}^{+6300}$ & 28000 & $_{-11000}^{+15000}$  & \ldots & \ldots & \ldots & \ldots & 11700 & $_{-4400}^{+9400}$  & 11790 & $_{-970}^{+740}$  & 9600 & $_{-230}^{+200}$ \\
IC~418    & 9000 & $_{-2300}^{+4800}$ & 15300 & $_{-6600}^{+16000}$ & 10400 & $_{-1800}^{+1700}$ & 4860 & $_{-850}^{+820}$ & 9800 & $_{-2000}^{+4300}$  & 9530 & $_{-370}^{+390}$  & 8780 & $_{-190}^{+150}$ \\
M~1-20    & 9400 & $_{-2700}^{+7100}$ & 9300 & $_{-3200}^{+5400}$ & 8700 & $_{-4700}^{+12000}$ & 10600 & $_{-5800}^{+6200}$ & 10900 & $_{-2900}^{+4100}$ & 11110 & $_{-610}^{+510}$ & 9760 & $_{-190}^{+200}$\\
NGC~40    & 1200 & $_{-220}^{+300}$  & 1840 & $_{-410}^{+440}$  & 890 & $\pm400$  & \ldots & \ldots   & 1320 & $_{-210}^{+200}$  & 8460 & $_{-220}^{+150}$  & 10390 & $\pm160$ \\
NGC~3242  & \ldots & \ldots  & 2350 & $_{-520}^{+690}$  & 1080 & $\pm420$  & 2100 & $_{-670}^{+760}$ & 1860 & $_{-350}^{+400}$  & 12170 & $_{-410}^{+470}$  & 11710 & $_{-230}^{+280}$ \\
NGC~3918  & \ldots & \ldots  & 5000 & $_{-1300}^{+1900}$  & 5480 & $_{-670}^{+1100}$  & 5920 & $_{-830}^{+1200}$ & 5570 & $_{-630}^{+970}$  & 10830 & $_{-320}^{+380}$  & 12540 & $_{-310}^{+290}$  \\
NGC~6826  & 1720 & $_{-310}^{+370}$  & 1980 & $_{-460}^{+540}$  & 1310 & $_{-420}^{+490}$  & 2140 & $_{-660}^{+710}$  & 1830 & $_{-250}^{+240}$  & 10400 & $_{-430}^{+530}$  & 9200 & $_{-120}^{+150}$ \\
[0.2ex] \hline \\
[-2.8ex]
\multicolumn{7}{l}{\ion{H}{ii} regions}\\
\hline  

M8  & 1420 & $_{-400}^{+560}$ & 1610 & $_{-180}^{+230}$ & 1600 & $_{-260}^{+210}$ & 3900 & $_{-2900}^{+5100}$& 1830 & $_{-440}^{+1300}$ & 8290 & $_{-150}^{+140}$ & 8040 & $_{-80}^{+100}$ \\
M16 & 1010 & $_{-190}^{+180}$ & 1420 & $_{-270}^{+290}$ & 1090 & $_{-490}^{+460}$ & \ldots & \ldots  & 1180 & $_{-200}^{+190}$ & 8310 & $_{-160}^{+140}$ & 7580 & $\pm130$ \\
M17 & 450 & $_{-80}^{+110}$ & 520 & $_{-140}^{+160}$ & 330 & $_{-220}^{+240}$ & 2100 & $_{-1300}^{+4600}$  & 490 & $_{-110}^{+210}$ & 8800 & $_{-220}^{+240}$ & 7960 & $_{-130}^{+100}$ \\
M20 & 230 & $\pm50$ & 350 &$\pm100$ & 390 & $_{-250}^{+300}$ & \ldots & \ldots  & 310 & $_{-60}^{+120}$ & 8250 & $_{-150}^{+160}$ & 7760 & $_{-60}^{+90}$ \\
M42 & 4930 & $_{-840}^{+1600}$ & 6100 & $_{-1800}^{+4700}$ & 5790 & $_{-390}^{+450}$ & 4860 & $_{-870}^{+990}$ & 5600 & $_{-680}^{+1200}$ & 10200 & $_{-230}^{+260}$ & 8290 & $_{-130}^{+120}$ \\
NGC~3576 & 1470 & $_{-240}^{+230}$ & 1330 & $_{-320}^{+390}$ & 2430 & $_{-690}^{+1200}$ & 3000 & $_{-1300}^{+1600}$ & 2090 & $_{-360}^{+570}$ & 8650 & $_{-160}^{+210}$ & 8430 & $\pm40$ \\
NGC~3603 & 2260 & $_{-480}^{+470}$ & 4000 & $_{-1000}^{+1600}$ & 3800 & $_{-1200}^{+1600}$ & 1700 & $_{-1100}^{+2100}$ & 3160 & $_{-510}^{+660}$ & 11120 & $_{-490}^{+460}$ & 9010 & $_{-130}^{+130}$ \\
[0.2ex] \hline\hline
\end{tabular}
\end{table*}
\renewcommand{\arraystretch}{1.}

\subsection{Ionic abundances}

The ionic abundances of \abariii, \abcliii, \abnii, \abneiii, \aboii, and \aboiii\ are computed using the
following CELs: [\ion{Ar}{iii}] $\lambda\lambda$7136, 7751, [\ion{Cl}{iii}] $\lambda\lambda$5518, 5538, 
[\ion{N}{ii}] $\lambda\lambda$6548, 6584, [\ion{Ne}{iii}] $\lambda\lambda$3869, 3968, 
[\ion{O}{ii}] $\lambda\lambda$3726, 3729, and [\ion{O}{iii}] $\lambda\lambda$4959,
5007, respectively. We do not consider the sulfur ions in our abundance determination because
of the still  unsolved ``sulfur problem'', whereby the values derived in many PNe are found to be
systematically lower than those in \ion{H}{ii} regions of the same metallicity \citep{hen04}. We
adopt \temp([\ion{N}{ii}]) for the calculations involving the single ionized ions and
\temp([\ion{O}{iii}]) for the double ionized ions; the adopted value of \dens\ is used in all cases. 

The \abariii\ abundance of DdDm~1 is calculated using the [\ion{Ar}{iii}] $\lambda\lambda$7135,
7751 line intensities of \citet{hen08} instead of those measured by \citet{wes05}, the reference
that we use for all the other line intensities for this object. The reason is that the intensities
provided by \citet{wes05} for these lines are too low when compared to measurements by other
authors \citep*{bar84, cle87, hen08}, which is unexpected since the angular diameter of this
object (0.6'') is smaller than the slit widths used for the observations. The intensities given by
\citet{wes05} imply $12+\log$(\abariii/H$^+$) = 4.99, whereas the values listed by
\citet{bar84, cle87, hen08} cover the range 5.50--5.69. 

We use \temp([\ion{O}{iii}]) and the average \dens\ to compute  the abundances of \abhei, \abheii,
\abcii, and \aboiii from RLs. We calculate He$^{+}$ as the average of the abundances derived from
the \ion{He}{i} lines $\lambda4471$, $\lambda5876$, and $\lambda6678$. The only exception is
DdDm~1, where we only use the first two lines because the abundance of \abhei\ derived from
$\lambda6678$ is too low. Since the intensity reported by \citet{wes05} for this line is also too
low when compared to the measurements of other authors \citep{bar84, cle87, hen08}, we think that
there is some problem with the intensities of \citet{wes05} for several lines in the red part of
the spectrum, such as [\ion{Ar}{iii}] $\lambda\lambda$7135,7751, \ion{He}{i} $\lambda6678$, and
[\ion{S}{ii}] $\lambda$6716. The He$^{++}$ abundances are derived using the \ion{He}{ii}
$\lambda$4686 line. We use the effective recombination coefficients of \citet{sto95} for \ion{H}{i} 
and \ion{He}{ii} and those computed by \citet{por12, por13} for \ion{He}{i}. 

The abundances of C$^{++}$ are computed using \ion{C}{ii} $\lambda4267$ and the case B effective 
recombination coefficients of \citet{dav00}. As for the \aboiii abundances, we use the total 
intensity of multiplet 1 of \ion{O}{ii}, after correcting for the contribution of undetected lines
with the formulae of \citet{pei05}, and the recombination coefficients of \citet{sto94}. 
We exclude from our calculations \ion{O}{ii} lines of multiplet 1 reported as blended in the 
papers from which we took the line intensities. In the case of NGC~40, there is only one detected 
\ion{O}{ii} feature from this multiplet and it is blended with a \ion{N}{iii} line, thus we prefer to use 
the abundance of O$^{++}$ derived by \citet{liu04b} from other multiplets. For the rest of the PNe, 
we do not consider other multiplets because their lines are weaker and likely suffer from blends. 
Indeed, we find that the O$^{++}$ abundances computed only from multiplet 1 are generally lower 
than those obtained using other multiplets.

The final ionic abundances derived from CELs and RLs and their associated uncertainties (see
Section~\ref{uncert}) are presented in Table~\ref{tab:4}. 

\renewcommand{\arraystretch}{1.2}
\begin{table*}
\caption{Ionic abundances from CELs and RLs: $X^{+i} = 12 + \log(X^{+i}/{\rm H}^+)$}             
\label{tab:4}      
\centering          
\begin{tabular}{lr@{}lr@{}lr@{}lr@{}lr@{}lr@{}lr@{}lr@{}lll}
\hline\hline   
\multicolumn{1}{c}{Object} & \multicolumn{2}{c}{Ar$^{++}$} & \multicolumn{2}{c}{Cl$^{++}$} & \multicolumn{2}{c}{N$^+$} & \multicolumn{2}{c}{Ne$^{++}$} & 
\multicolumn{2}{c}{O$^+$} & \multicolumn{2}{c}{O$^{++}$} & 
\multicolumn{2}{c}{He$^+$} & \multicolumn{2}{c}{He$^{++}$} & \multicolumn{1}{c}{C$^{++}$} & \multicolumn{1}{c}{O$^{++}$}\\
\multicolumn{1}{c}{} & \multicolumn{2}{c}{CELs} & \multicolumn{2}{c}{CELs} & \multicolumn{2}{c}{CELs} & \multicolumn{2}{c}{CELs} & \multicolumn{2}{c}{CELs} & \multicolumn{2}{c}{CELs} & \multicolumn{2}{c}{RLs} &
\multicolumn{2}{c}{RLs} & \multicolumn{1}{c}{RLs} & \multicolumn{1}{c}{RLs}\\
\hline 
\multicolumn{17}{l}{O-rich dust PNe}\\    
\hline                
Cn~1-5        & 6.38 & $\pm0.03$ & 5.22 & $\pm0.05$ & 8.19 & $_{-0.03}^{+0.07}$  & 8.28 & $\pm0.04$ & 8.26 & $_{-0.04}^{+0.16}$ & 8.70 & $\pm0.04$  & 11.01 & $\pm0.01$ & \ldots & \ldots & 9.09 & 9.01  \\
DdDm~1      & 5.61 & $\pm0.04$ & 4.59 & $\pm0.04$ & 6.75 & $_{-0.04}^{+0.08}$  & 7.19 & $\pm0.06$ & 7.38 & $_{-0.09}^{+0.20}$ & 7.87 & $\pm0.04$ & 10.94 & $\pm0.02$ & \ldots & \ldots & \ldots & 8.39 \\
H~1-50        & 5.99 & $\pm0.03$ & 4.88 & $\pm0.04$ & 7.01 & $_{-0.03}^{+0.06}$   & 7.94 & $\pm0.04$  & 7.38 & $_{-0.05}^{+0.13}$ & 8.64 & $\pm0.05$ & 11.00 & $\pm0.01$ & 10.04 & $\pm0.02$ & 8.35 & 9.09  \\
M~1-42        & 6.27 & $\pm0.03$ & 5.04 & $\pm0.04$ & 7.89 & $\pm0.04$               & 8.03 & $\pm0.04$ & 7.64 & $_{-0.04}^{+0.07}$ & 8.39 & $\pm0.04$ & 11.20 & $\pm0.01$ & 10.04 & $\pm0.02$ & 9.39 & 9.63 \\
M~2-27       & 6.50 & $\pm0.03$ & 5.29 & $_{-0.02}^{+0.06}$ & 7.66 & $\pm0.07$   & 8.37 & $\pm0.04$  & 7.64 & $_{-0.10}^{+0.17}$ & 8.84 & $\pm0.04$ & 11.15 & $\pm0.01$ & 8.84 & $\pm0.02$ & 8.85 & 9.27 \\
M~2-31       & 6.15 & $\pm0.04$ & 5.00 & $\pm0.04$ & 7.04 & $\pm0.05$               & 8.06 & $\pm0.04$ & 7.24 & $_{-0.06}^{+0.12}$ & 8.64 & $\pm0.05$ & 11.07 & $\pm0.01$ &\ldots & \ldots & \ldots & \ldots \\
M~2-42       & 6.16 & $\pm0.03$ & 5.14 & $\pm0.07$ & 7.00 & $_{-0.03}^{+0.07}$   & 8.04 & $\pm0.04$ & 7.33 & $_{-0.06}^{+0.17}$ & 8.73 & $\pm0.04$ & 11.06 & $\pm0.01$ & 8.45 & $\pm0.08$ &  \ldots & \ldots  \\
MyCn~18    & 6.33 & $\pm0.03$ & 5.31 & $\pm0.03$ & 7.57 & $\pm0.04$  & 8.18 & $\pm0.05$ & 7.68 & $_{-0.06}^{+0.09}$ & 8.53 & $\pm0.04$ & 11.00 & $\pm0.02$ & 8.66 & $\pm0.02$ &  8.36 & 8.87 \\
NGC~3132 & 6.38 & $\pm0.03$ & 5.18 & $\pm0.03$ & 8.19 & $\pm0.04$                & 8.30 & $\pm0.04$ &  8.47 & $\pm0.06$ & 8.53 & $\pm0.05$ & 11.05 & $\pm0.01$ & 9.52 & $\pm0.02$ & 8.83 & 8.81 \\
NGC~6210 & 5.93 & $\pm0.04$ & 4.70 & $\pm0.05$ & 6.47 & $\pm0.05$                & 8.12 & $\pm0.06$ & 7.09 & $\pm0.10$ & 8.55 & $\pm0.06$ & 11.02 & $\pm0.02$ & 9.28 & $\pm0.02$ & 8.80 & 8.97 \\
NGC~6439  & 6.35 & $\pm0.03$ & 5.12 & $\pm0.03$ & 7.52 & $\pm0.05$               & 8.16 & $\pm0.04$ & 7.67 & $\pm0.07$ & 8.60 & $\pm0.04$ & 11.05 & $\pm0.02$ & 10.31 & $\pm0.02$ & 8.99 & 9.06 \\
NGC~6543  & 6.48 & $\pm0.03$ & 5.22 & $\pm0.06$ & 6.80 & $_{-0.05}^{+0.10}$  & 8.24 & $\pm0.05$ & 7.20 & $_{-0.06}^{+0.20}$ & 8.79 & $\pm0.05$ & 11.06 & $\pm0.01$ & \ldots & \ldots & 8.77 & 9.07  \\
NGC~7026  & 6.22 & $\pm0.02$ & 5.16 & $_{-0.02}^{+0.06}$ & 7.56 & $_{-0.04}^{+0.08}$  & 8.19 & $\pm0.03$ & 7.66 & $_{-0.06}^{+0.22}$ & 8.64 & $\pm0.03$ & 11.03 & $\pm0.01$ & 10.11 & $\pm0.02$ & 8.94 & 9.07 \\
[0.2ex] \hline \\
[-2.8ex]
\multicolumn{17}{l}{C-rich dust PNe}\\
\hline  
Hu~2-1         & 5.69 & $\pm0.03$ & 4.51 & $_{-0.03}^{+0.07}$ & 6.93 & $_{-0.04}^{+0.16}$  & 7.44 & $\pm0.04$ & 7.50 & $_{-0.11}^{+0.42}$ & 8.23 & $_{-0.04}^{+0.07}$   & 10.91 & $\pm0.01$ & 8.39 & $\pm0.02$ &  8.62 & 8.71  \\
IC~418         & 6.00 & $\pm0.03$ & 4.82 & $\pm0.04$ & 7.63 & $\pm0.06$  & 6.78 & $\pm0.04$ & 8.34 & $_{-0.15}^{+0.11}$ & 8.09 & $\pm0.04$ & 10.96 & $\pm0.01$ & \ldots & \ldots & 8.74 & 8.24  \\
M~1-20        & 5.77 & $\pm0.03$ & 4.65 & $_{-0.06}^{+0.12}$ & 6.79 & $_{-0.05}^{+0.11}$  & 7.69 & $\pm0.04$ & 7.34 & $_{-0.10}^{+0.28}$ & 8.55 & $\pm0.05$ & 10.99 & $\pm0.01$ & 7.60 & $_{-0.16}^{+0.10}$ & 8.67 & 8.66  \\
NGC~40      & 5.67 & $\pm0.02$ & 4.65 & $\pm0.03$ & 7.94 & $\pm0.04$ & 5.80 &$\pm0.03$ & 8.65 & $\pm0.06$ & 7.11 & $\pm0.03$ & 10.78 & $\pm0.01$ & 7.56 & $_{-0.16}^{+0.13}$ & 8.81 & 8.33  \\
NGC~3242 & 5.60 & $\pm0.03$ & 4.40 & $\pm0.03$ & 5.55 & $\pm0.04$  & 7.88 & $\pm0.04$ & 6.50 & $\pm0.06$ & 8.43 & $\pm0.05$ & 10.87 & $\pm0.02$ & 10.34 & $\pm0.02$ & 8.79 & 8.84 \\
NGC~3918 & 5.93 & $\pm0.03$ & 4.71 & $\pm0.04$ & 7.15 & $\pm0.05$  & 7.88 & $\pm0.04$ & 7.66 & $_{-0.06}^{+0.09}$ & 8.45 & $\pm0.05$ & 10.82 & $\pm0.02$ & 10.55 & $\pm0.02$ & 8.70 & 8.82  \\
NGC~6826  & 6.06 & $\pm0.02$ & 4.85 & $\pm0.03$ & 6.22 & $\pm0.06$  & 7.88 & $\pm0.03$ & 7.00 & $\pm0.08$ & 8.55 & $\pm0.04$ & 11.00 & $\pm0.01$ & 7.34 & $_{-0.16}^{+0.10}$ & 8.74 & 8.82 \\
[0.2ex] \hline \\
[-2.8ex]
\multicolumn{17}{l}{\ion{H}{ii} regions}\\
\hline  
M8            & 6.16 & $\pm0.02$  & 5.04 &  $\pm0.03$ & 7.54 &  $\pm0.03$               & 7.05 & $\pm0.04$ & 8.37 &  $_{-0.04}^{+0.12}$ & 7.89 &  $\pm0.03$ & 10.84 & $\pm0.01$ & \ldots & \ldots & 8.31 & 8.24  \\
M16          & 6.19 &  $\pm0.04$ & 5.09 &  $\pm0.04$ & 7.73 &  $\pm0.03$  & 7.15 & $\pm0.05$ & 8.47 &  $\pm0.05$ & 7.92 &  $\pm0.04$ & 10.90 & $\pm0.01$ &  \ldots & \ldots & 8.39 & 8.31\\
M17          & 6.30 & $\pm0.03$  & 5.06 &  $\pm0.03$ & 6.84 &  $\pm0.04$               & 7.76 & $\pm0.03$ & 7.80 &  $_{-0.04}^{+0.07}$ & 8.45 &  $\pm0.03$ & 10.98 & $\pm0.01$ & \ldots & \ldots & 8.73 & 8.72\\
M20          & 6.16 & $\pm0.02$  & 5.04 &  $\pm0.03$ & 7.59 &  $\pm0.04$               & 6.68 & $\pm0.03$ & 8.48 &  $\pm0.05$ & 7.73 &  $\pm0.02$ & 10.85 & $\pm0.01$ & \ldots & \ldots &  8.19 & 8.18\\
M42          & 6.33 & $\pm0.03$  & 5.12 &  $\pm0.04$ & 6.87 &  $\pm0.04$               & 7.77 & $\pm0.03$ & 7.71 &  $_{-0.05}^{+0.10}$ & 8.43 &  $\pm0.05$  & 10.98 & $\pm0.01$ & \ldots & \ldots & 8.35 & 8.61\\
NGC~3576 & 6.30 & $\pm0.05$  & 4.94 & $\pm0.03$  & 7.06 &  $\pm0.04$             & 7.70 & $\pm0.02$ & 8.11 &  $\pm0.06$              & 8.37 &  $\pm0.01$ & 10.97 &$\pm0.01$ & \ldots & \ldots & 8.44 & 8.67 \\
NGC~3603 & 6.29 & $\pm0.02$  & 5.04 &  $\pm0.04$ & 6.48 &  $\pm0.06$             & 7.78 & $\pm0.03$ & 7.34 &  $_{-0.06}^{+0.12}$ & 8.44 &  $\pm0.03$ & 11.00 & $\pm0.01$ &  \ldots & \ldots & 8.49 & 8.71 \\
[0.2ex] \hline\hline
\end{tabular}
\end{table*}
\renewcommand{\arraystretch}{1.}

\subsection{Total abundances}

To obtain total elemental abundances one needs to take into account the contribution of unobserved
ions by using ICFs. Many studies still use ICFs based on similarities of ionization potentials. Those
however ignore the fact that a small difference in the ionization potential may lead to a significant
change in the ionization structure \citep[e.g.][]{sim08,del14b}, and that the ionization structure is
not only governed by the spectral energy distribution of the stellar radiation field but also by the
importance of recombination and charge exchange reactions. Photoionization models allow a much better
estimate of ICFs. For PNe, the ICFs proposed by \citet{kin94} have been  used for two decades.
However, there is no real  documentation on how they were obtained and there is no estimate of their
uncertainties. A set of formulae has been proposed recently by \citet{del14b}, based on an extensive
grid of  \textit{ab initio} photoionization models aiming at representing the whole manifold of PNe. 
The large set of considered photoionization models allowed, for the 
first time, an estimation of the error bars on ICFs. They are computed using the analytical expressions 
provided by \citet{del14b}, which are based on the maximum dispersion of each ICF obtained from 
the grid of photoionization models. Those ICFs were tested by applying them to a large sample of PNe and
checking that the derived abundance ratios did not depend on the degree of ionization of the objects.

The sample of PNe considered here is a specific sample excluding high density objects and objects 
ionized by very hot stars. We therefore checked if the ICFs from \citet{del14b}, which were obtained by 
considering a wider sample of physical conditions, remain valid for the restricted subsample of 
photoionization models better representing our observational sample, i.e., we discard the models with 
very high gas density or with ionizing stars of very high effective temperatures. We found that the ICFs 
presented in Delgado-Inglada (2014)  remain valid for the restricted subsample of models, both as 
regards the analytical formulae for the ICFs and as regards the error estimates.

We thus adopted the ICFs derived by \citet{del14b} to calculate the abundance ratios Ar/O, C/O, Cl/O,
Ne/O, He/H, and O/H. The ICF for O is based on He$^{++}$/(He$^+$+He$^{++}$), the ICFs for Ar, C, Cl,
and He are based on O$^{++}$/(O$^+$+O$^{++}$), and the ICF for Ne is based on both ratios. As for N,
we show here the results obtained with the ICF derived by \citet{del14b} as well as those derived
with the widely used ICF, N/O = N$^+$/O$^+$ (which are the ones listed in Table~\ref{tab:5}). We will
discuss the differences between both estimates of N/O in section~4.5.

For the \ion{H}{ii} regions we are considering in this paper, the situation is a priori worse.
Formulae for ICFs  based on photoionization models have been proposed  by \citet{izo06}, but they were
proposed and  tested for giant extragalactic \ion{H}{ii} regions, and there is a priori no reason why
they should apply to slit observations of nearby dense \ion{H}{ii} regions, some of them ionized by
single stars. In fact, we find that they introduce a dependence of the total abundances of argon and
neon on the degree of ionization for the \ion{H}{ii} regions in our sample. In absence of any adequate
set of  ICFs, we used for our \ion{H}{ii} regions the same expressions that we used for our PNe. In
principle, we could  trim the grid of models built for PNe to retain only those that would be adequate
for the \ion{H}{ii} regions we  consider (ie. remove the models with effective temperatures above
50000 K and all those that have a shell structure or are density bounded) and use the trimmed grid to
compute ICFs that would be relevant for  our sample of \ion{H}{ii} regions. In practise, we checked
that the formulae given by \citet{del14b} actually fit the trimmed grid quite well (except for the
determination of Ar/O in objects where O$^{++}$/(O$^+$+O$^{++}$) is larger than 0.5, where the 
resulting bias on Ar/O can go up to $\pm$0.09 dex.

One important advantage of the ICFs used here to determine the Cl and Ar abundances is that they can
be used when only lines of Cl$^{++}$ and Ar$^{++}$, respectively, are observed. These ions can  be
analysed using relatively bright lines in the optical spectrum, which makes it easier to achieve a
homogeneous calculation of the total abundances. Other ICFs in the literature include more ions that
have relatively weak emission lines, such as the [\ion{Ar}{iv}] $\lambda\lambda4711,4740$ lines in
some objects, or that lie in a spectral range that it is not often observed, such as [\ion{Cl}{ii}]
$\lambda9122$. Moreover, we found, from the grid of photoionization models computed in 
\citet{del14b}, that the other ICFs in the literature for Ar and Cl introduce a higher dispersion  in
the derived abundances.

As for the O/H values computed from RLs, for the \ion{H}{ii} regions we adopt those provided by
\citet{est05}. They are based on the O$^{+}$ and O$^{++}$ abundances derived with \ion{O}{i} and
\ion{O}{ii} RLs for some objects; for other objects \ion{O}{i} RLs could not be measured properly and
the O$^{+}$ abundances are based on the abundances implied by CELs corrected upwards to account for
the possible presence of small temperature fluctuations within the nebulae. We note that the results
of \citet{est05} show that some of the dispersion in these values of O/H (and in the corresponding
values based on CELs) arises from the Galactic abundance gradient. The assumption of small 
temperature fluctuations might not be right for the PNe \citep{wes05}. Besides, only IC~418
has measurements of suitable \ion{O}{i} RLs \citep{sha03}. Hence, we derived their RLs
oxygen abundances using the O$^{++}$ abundances implied by multiplet 1 of \ion{O}{ii} and the
ionization fraction for this ion inferred from the CELs analysis. 

The final abundances for each PN and \ion{H}{ii} region are listed in Table~\ref{tab:5}.

\renewcommand{\arraystretch}{1.2}
\begin{table*}
\caption{Abundance ratios: $\{X/{\rm H}\} = 12 + \log(X/{\rm H})$}             
\label{tab:5}      
\centering          
\begin{tabular}{lr@{}lr@{}lr@{}lr@{}lr@{}lr@{}lr@{}lrc}
\hline\hline   
\multicolumn{1}{c}{Object} & \multicolumn{2}{c}{\{Ar/H\}} &  \multicolumn{2}{c}{\{Cl/H\}} & \multicolumn{2}{c}{\{N/H\}$^*$} & \multicolumn{2}{c}{\{Ne/H\}} 
& \multicolumn{2}{c}{\{O/H\}} & \multicolumn{2}{c}{\{He/H\}} & \multicolumn{2}{c}{$\log$(C/O)} & \multicolumn{1}{c}{\{O/H\}} & \multicolumn{1}{c}{Galactic}\\
\multicolumn{1}{c}{} & \multicolumn{2}{c}{CELs} & \multicolumn{2}{c}{CELs} & \multicolumn{2}{c}{CELs} & \multicolumn{2}{c}{CELs} & 
\multicolumn{2}{c}{CELs} & \multicolumn{2}{c}{RLs} & \multicolumn{2}{c}{RLs} & \multicolumn{1}{c}{RLs} & \multicolumn{1}{c}{component}\\
\hline 
\multicolumn{17}{l}{O-rich dust PNe}\\    
\hline  
Cn~1-5        & 6.45 & $_{-0.27}^{+0.13}$  & 5.35 & $\pm0.08$  & 8.77 & $_{-0.24}^{+0.12}$  & 8.66 & $_{-0.04}^{+0.09}$ & 8.84 & $_{-0.02}^{+0.06}$ & 11.10 & $\pm0.01$ & 0.02 & $_{-0.38}^{+0.33}$ & 9.14 & bulge \\
DdDm~1      & 5.69 & $_{-0.32}^{+0.14}$  & 4.73 & $_{-0.10}^{+0.04}$  & 7.36 & $_{-0.25}^{+0.11}$   & 7.55 &  $\pm0.07$ & 7.99 & $_{-0.03}^{+0.07}$    & 10.94 & $\pm0.02$  & \ldots & \ldots & 8.51 & halo \\
H~1-50         & 6.22 & $_{-0.24}^{+0.47}$  & 5.18 & $_{-0.12}^{+0.18}$  & 8.32 & $_{-0.27}^{+0.11}$  & 7.96 & $_{-0.07}^{+0.03}$  & 8.69 & $\pm0.04$   & 11.04 & $\pm0.01$ & $-0.68$ & $\pm0.38$ & 9.14 & bulge \\
M~1-42        & 6.42 & $_{-0.28}^{+0.15}$  & 5.23 & $_{-0.10}^{+0.03}$  & 8.73 & $_{-0.22}^{+0.15}$  & 8.05 &  $\pm0.07$ & 8.48 & $\pm0.04$  & 11.22 & $\pm0.01$ & $-0.24$ & $_{-0.40}^{+0.36}$ & 9.72 & bulge \\
M~2-27        & 6.69 & $_{-0.23}^{+0.19}$  & 5.56 & $_{-0.09}^{+0.11}$  & 8.89 & $_{-0.23}^{+0.14}$  & 8.54 &  $\pm0.06$ & 8.87 & $_{-0.03}^{+0.06}$  & 11.16 & $\pm0.01$ & $-0.37$ & $\pm0.38$ & 9.30 & bulge \\
M~2-31        & 6.36 & $_{-0.08}^{+0.58}$  & 5.31 & $_{-0.10}^{+0.26}$  & 8.46 & $_{-0.24}^{+0.15}$  & 8.20 &  $\pm0.06$ & 8.66 & $\pm0.04$ & 11.07 & $\pm0.01$ & \ldots &  \ldots & \ldots & bulge \\
M~2-42        & 6.37 & $_{-0.16}^{+0.56}$  & 5.45 & $_{-0.12}^{+0.24}$  & 8.41 & $_{-0.23}^{+0.15}$  & 8.18 &  $\pm0.05$ & 8.74 & $\pm0.04$  & 11.06 & $\pm0.01$ & \ldots &  \ldots & 9.50 & bulge \\
MyCn~18     & 6.47 & $_{-0.32}^{+0.15}$  & 5.50 & $_{-0.11}^{+0.03}$  & 8.48 & $_{-0.24}^{+0.14}$  & 8.41 &  $\pm0.06$ & 8.58 & $\pm0.04$  & 11.00 & $\pm0.01$ & $-0.49$ & $\pm0.37$ & 8.93 & disc \\
NGC~3132  & 6.40 & $_{-0.27}^{+0.14}$  & 5.30 & $_{-0.10}^{+0.04}$  & 8.53 & $_{-0.20}^{+0.14}$  & 8.36 &  $\pm0.06$ & 8.81 & $\pm0.03$  & 11.06 & $\pm0.01$ & $-0.13$ & $\pm0.34$ & 9.08 & disc \\
NGC~6210  & 6.15 & $_{-0.10}^{+0.59}$  & 5.02 & $_{-0.06}^{+0.26}$  & 7.95 & $_{-0.24}^{+0.15}$  & 8.16 &  $\pm0.07$ & 8.57 & $\pm0.05$  & 11.02 & $\pm0.02$ & $-0.10$ & $_{-0.33}^{+0.39}$ & 8.99 & disc \\
NGC~6439  & 6.55 & $_{-0.25}^{+0.14}$  & 5.37 & $_{-0.10}^{+0.04}$  & 8.54 & $_{-0.27}^{+0.15}$  & 8.20 &  $\pm0.05$ & 8.69 & $\pm0.05$  & 11.12 & $\pm0.02$ &  $-0.05$ & $\pm$ & 9.15 & bulge \\
NGC~6543  & 6.70 & $_{-0.08}^{+0.60}$  & 5.57 & $_{-0.11}^{+0.26}$  & 8.40 & $_{-0.25}^{+0.13}$  & 8.36 &  $\pm0.07$ & 8.80 & $_{-0.03}^{+0.06}$  & 11.06 & $\pm0.02$ & $-0.23$ & $_{-0.36}^{+0.40}$ & 9.08 & disc \\
NGC~7026  & 6.42 & $_{-0.34}^{+0.12}$  & 5.41 & $_{-0.09}^{+0.05}$  & 8.62 & $_{-0.31}^{+0.13}$  & 8.22 &  $\pm0.05$ & 8.71 & $_{-0.02}^{+0.06}$  & 11.08 & $\pm0.01$  & $-0.10$ & $_{-0.38}^{+0.34}$ & 9.14 & disc \\
[0.2ex] \hline \\
[-2.8ex]
\multicolumn{17}{l}{C-rich dust PNe}\\
\hline  
Hu~2-1        & 5.81 & $_{-0.31}^{+0.14}$ & 4.68 & $_{-0.10}^{+0.05}$ & 7.73 & $\pm0.16$ & 7.71 & $_{-0.04}^{+0.20}$ & 8.30 & $_{-0.03}^{+0.14}$ & 10.91 & $\pm0.01$ & $-0.09$ & $_{-0.39}^{+0.32}$ & 8.78 & disc\\
IC~418         & 6.05 & $_{-0.32}^{+0.16}$ & 4.94 & $_{-0.11}^{+0.04}$ & 7.82 & $_{-0.11}^{+0.16}$ & 7.54 & $_{-0.12}^{+0.05}$ & 8.54 & $\pm0.09$ & 10.96 & $\pm0.01$ & 0.25 & $_{-0.34}^{+0.31}$ & 8.69 & disc\\
M~1-20        & 5.96 & $_{-0.27}^{+0.21}$ & 4.92 & $_{-0.12}^{+0.11}$ & 8.03 & $_{-0.16}^{+0.13}$ & 7.85 & $_{-0.05}^{+0.11}$ & 8.58 & $_{-0.03}^{+0.06}$ & 10.99 & $\pm0.01$ & 0.06 & $_{-0.37}^{+0.33}$ & 8.68 & bulge\\
NGC~40      & 6.17 & $_{-0.63}^{+0.14}$ & 4.92 & $_{-0.10}^{+0.05}$ & 7.96 & $_{-0.11}^{+0.15}$ & 7.60 & $_{-0.15}^{+0.52}$ & 8.66 & $\pm0.05$ & $>$10.78 &  & $-0.48$ & $_{-0.56}^{+0.33}$ & 9.89 & disc\\
NGC~3242 & 5.91 & $_{-0.06}^{+0.61}$& 4.89 & $_{-0.12}^{+0.25}$ & 7.56 & $_{-0.12}^{+0.15}$   & 7.95 & $_{-0.08}^{+0.04}$ & 8.50 & $\pm0.06$  & 10.98 & $\pm0.01$ & 0.03 & $_{-0.34}^{+0.45}$ & 8.91 & disc\\
NGC~3918 & 6.18 & $_{-0.36}^{+0.15}$ & 5.01 & $_{-0.10}^{+0.05}$ & 8.13 & $\pm0.14$  & 8.00 & $\pm0.08$ & 8.63 & $\pm0.06$ & 11.01 & $\pm0.01$ & $-0.12$ & $\pm0.36$ & 9.01 & disc\\
NGC~6826  & 6.28 & $_{-0.05}^{+0.60}$ & 5.19 & $_{-0.09}^{+0.26}$ & 7.78 & $_{-0.10}^{+0.16}$ & 8.01 & $\pm0.06$ & 8.56 & $_{-0.02}^{+0.03}$ & 11.00 & $\pm0.01$ & $-0.01$ & $_{-0.36}^{+0.39}$ & 8.83 & disc\\
[0.2ex] \hline \\
[-2.8ex]
\multicolumn{17}{l}{\ion{H}{ii} regions}\\
\hline  
M8   &            6.25 & $_{-0.31}^{+0.16}$ & 5.18 & $_{-0.09}^{+0.06}$ & 7.67 & $_{-0.11}^{+0.16}$   & 7.65 &  $_{-0.05}^{+0.09}$ & 8.50 & $_{-0.03}^{+0.10}$  & $>$10.84 &  & $-0.27$ & $\pm0.30$ & 8.71 & disc  \\
M16 &            6.30 & $_{-0.25}^{+0.14}$ & 5.24 & $_{-0.11}^{+0.05}$ & 7.84 & $_{-0.11}^{+0.18}$   & 7.81 &  $\pm0.05$ & 8.58 & $\pm0.04$  & $>$10.90 &  & $-0.29$ & $\pm0.33$ & 8.80 & disc   \\
M17 &            6.41 & $_{-0.34}^{+0.15}$ & 5.22 & $_{-0.09}^{+0.05}$ & 7.58 & $\pm0.13$   & 7.85 & $\pm0.05$ & 8.54 & $\pm0.03$  & 10.98 & $\pm0.01$ & $-0.01$ & $_{-0.33}^{+0.36}$ & 8.76 & disc  \\
M20 &            6.32 & $_{-0.31}^{+0.15}$ & 5.21 & $_{-0.10}^{+0.04}$ & 7.66 & $_{-0.11}^{+0.15}$   & 7.50 & $\pm0.06$ & 8.55 & $\pm0.04$  & $>$10.85 &  & $-0.47$ & $_{-0.29}^{+0.34}$  & 8.71 & disc   \\
M42 &            6.46 & $_{-0.29}^{+0.16}$ & 5.29 & $_{-0.10}^{+0.04}$ & 7.68 & $_{-0.10}^{+0.15}$   & 7.84 & $\pm0.05$ & 8.51 & $\pm0.05$  & 10.98 & $\pm0.01$ & $-0.26$ & $_{-0.33}^{+0.36}$ & 8.65 & disc  \\
NGC~3576 &   6.34 & $_{-0.30}^{+0.13}$ & 5.06 & $_{-0.10}^{+0.04}$ & 7.51 & $_{-0.08}^{+0.15}$ & 7.89 & $\pm0.05$ & 8.56 & $\pm0.03$  & 10.97 & $\pm0.01$ & $-0.33$ & $\pm0.34$ & 8.74 & disc   \\
NGC~3603 &   6.47 & $_{-0.30}^{+0.14}$ & 5.28 & $_{-0.11}^{+0.05}$ & 7.60 & $\pm0.13$ & 7.82 & $\pm0.05$ & 8.47 & $\pm0.03$  & 11.00 & $\pm0.01$ & $0.20$ & $_{-0.30}^{+0.34}$  & 8.72 & disc  \\
[0.2ex] \hline \\
[-2.8ex]
\multicolumn{17}{l}{
\begin{minipage}{8.0cm}
\small $^{*}$ N/H computed from N/O = N$^+$/O$^+$.
\end{minipage}
}\\ 
\end{tabular}
\end{table*}
\renewcommand{\arraystretch}{1.}

\subsection{Uncertainties}\label{uncert}

The uncertainties in the physical conditions and the ionic abundances are computed 
through a Monte Carlo simulation. We generate 200 random values for each line intensity 
using a Gaussian distribution centred in the observed line intensity and with a sigma 
equal to the associated uncertainty. For higher number of Monte Carlo simulations, 
the errors in the computed quantities remain the same.

We derive the values of \temp, \dens, and the ionic abundances for every Monte Carlo 
run and calculate the errors associated with each quantity from the 16 and 84 percentiles; 
these values define a confidence interval of 68 per cent which is equivalent to one standard 
deviation in a Gaussian distribution. We assign a typical uncertainty of 5\% and 15\% to 
the C$^{++}$ and O$^{++}$ ionic abundances, respectively, computed from RLs. We use this 
simplified approach because the determination of uncertainties for the O$^{++}$ abundances 
implied by RLs is difficult and unreliable, and the C$^{++}$ abundances are only used to 
calculate the C/O abundance ratio from the ratio C$^{++}$/O$^{++}$. Some of the \ion{O}{ii} 
lines from multiplet 1 are not observed and we have to estimate their contribution to the multiplet, 
which can depend on the electron density. In addition, some of the \ion{O}{ii} lines could be 
blended with nearby weaker lines. An uncertainty of 15\% corresponds to the typical difference 
between the abundances derived from individual lines of multiplet 1 and from the total intensity 
of the multiplet and also to the differences with the O$^{++}$ computed with other multiplets 
\citep{gar13, liu04b}. A complete analysis of the uncertainties related to the O$^{++}$ abundances 
implied by RLs should also take into account the information provided by the other multiplets and 
would be rather onerous.  We consider that our simple approach is sufficient for the purposes 
of this paper. The uncertainty of 5\% for the C$^{++}$ abundances 
arises from the typical uncertainty of the intensity of \ion{C}{ii} 4267 relative to H$\beta$ in our 
sample objects.

As for the total abundances, we consider not only the uncertainties associated with the physical 
conditions and the ionic abundances but also those related to the adopted ICF. To estimate the
uncertainties associated with the ICFs we perform another Monte Carlo simulation. We construct a
uniform distribution for each ICF, where the central value is the ICF calculated from the observed
degree of ionization, and the minimum and maximum values are obtained from the uncertainties
associated with each ICF provided in \citet{del14b}. From this distribution we generate random
values that we use to calculate the total abundances for the 200 runs. The errors associated with
the abundance of each element are given by the 16 and 84 percentiles. Our computed uncertainties in
the total abundances are probably overestimated since we are using the maximum dispersion in each
ICF, created by a large grid of photoionization models, whereas real nebulae are probably better
represented by a smaller set of models.

In the case of nitrogen, we use the method described above when using the ICF from \citet{del14b},
and we adopt a uniform distribution of errors of $\pm0.2$ dex when using N/O = N$^+$/O$^+$, which 
seems reasonable given the dispersion we get for abundance ratios like N/O and N/Cl in the 
sample \ion{H}{ii} regions.

As for He, \citet{del14b} proposed that one should compute the total He abundance by adding the
contributions of He$^+$ and  He$^{++}$, including the contribution of He$^0$ in the uncertainties
associated with He/H. There are four objects in the sample (three \ion{H}{ii} regions and one PN)
with a very low degree of ionization, where the abundance of He$^0$ is important and thus, the value
of He/H is very uncertain. We represent their He abundance in the figures as a lower limit and we only
plot the errors associated with the He ionic abundances and not with the ICF. 

Each of the ICFs derived in \citet{del14b} is valid for a certain range in the degree of ionization,
given by O$^{++}$/(O$^+$+O$^{++}$) or He$^{++}$/(He$^+$+He$^{++}$); outside this range the ICF 
formula is not necessarily correct. Here we use the ICFs from \citet{del14b} in all the objects,
regardless of their degree of ionization, but we  assign higher uncertainties in the ICFs when the
objects fall outside the range of validity. The low dispersion found in some of the abundance ratios,
such as Ar/Cl, suggests that the adopted ICFs are still valid outside the ranges proposed in
\citet{del14b}, at least in some cases.

Since the ICFs for many elements are based on the O$^{++}$/(O$^+$+O$^{++}$) ratio, 
the uncertainties introduced by these corrections are not independent. If we do not consider 
this effect, we get very large uncertainties for some of the abundance ratios considered below, 
namely, Ar/Cl and Ne/Cl. We have used the model results of \citet{del14b} to obtain better 
estimates of the uncertainties in those cases. The uncertainties in $\log$(Ar/Cl) associated with 
the ICF are: 
\begin{eqnarray*}
+0.30/-0.90\: {\rm dex}, {\rm for\:} {\rm O}^{++}/({\rm O}^{+}+{\rm O}^{++}) <0.05 \nonumber\\
+0.20/-0.15\: {\rm dex}, {\rm for\:} 0.05\leq{\rm O}^{++}/({\rm O}^{+}+{\rm O}^{++}) <0.45 \nonumber\\
+0.10/-0.04\: {\rm dex}, {\rm for\:} 0.45\leq{\rm O}^{++}/({\rm O}^{+}+{\rm O}^{++}) \leq0.95 \nonumber\\
+0.66/-0.01\: {\rm dex}, {\rm for\:} {\rm O}^{++}/({\rm O}^{+}+{\rm O}^{++}) >0.95. \nonumber
\end{eqnarray*}
As for the $\log$(Ne/Cl) values, the uncertainties associated with the ICF are: 
\begin{eqnarray*}
+0.70/-1.30\: {\rm dex}, {\rm for\:} {\rm O}^{++}/({\rm O}^{+}+{\rm O}^{++}) <0.1 \nonumber\\
+0.13/-0.08\: {\rm dex}, {\rm for\:} 0.1\leq{\rm O}^{++}/({\rm O}^{+}+{\rm O}^{++}) \leq0.95 \nonumber\\
+0.20/-0.30\: {\rm dex}, {\rm for\:} {\rm O}^{++}/({\rm O}^{+}+{\rm O}^{++}) >0.95. \nonumber
\end{eqnarray*}

We do not include the uncertainties on the atomic data since it is very difficult to quantify them 
but we have selected the atomic data that we consider more reliable \citep[see e.g.][]{sta13}.

The final adopted value for each parameter is the value computed from the observed intensities 
except for the electron density, for which we prefer the median of the distribution of average 
densities obtained from the available diagnostics in each run. In the low- and high-density limits, 
the median is less affected by the large variations obtained in the density values due to small 
variations in the intensity ratios. Note that the electron density is a special case; for some 
of the other calculated quantities, the use of the median would lead to similar results, for other 
quantities, such as the total abundances, it is not a good idea to use the median since their 
distributions depend on the distribution of the uncertainties due to the ICFs, for which we can 
only estimate the width, not the shape.

\section{Results and discussion}

\subsection{Chlorine as a proxy for metallicity}

\begin{figure}
\centering
\includegraphics[width=\hsize, trim = 30 15 40 0, clip =yes]{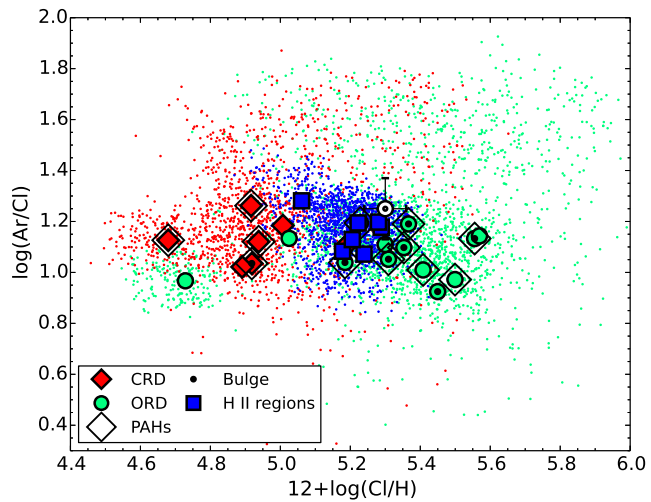}
\caption{Values of Ar/Cl as a function of Cl/H. The green circles represent ORD PNe, the red diamonds
CRD PNe, and the blue squares the \ion{H}{ii} regions. We mark with open diamonds the PNe with PAH
emission and with a small black dot the PNe from the bulge. The coloured small dots correspond to the
values obtained for the 200 Monte Carlo runs performed for each object. The protosolar abundances of
\citet{lod10} are overplotted with the solar symbol. The CRD PN NGC~6826, with
$12+\log(\mbox{Cl/H})=5.19$ and $\log(\mbox{Ar/Cl})=1.09$, is hidden behind the \ion{H}{ii} regions.
\label{fig:ArCl}}
\end{figure}

We have derived the abundances of He, C, N, O, Ne, Cl, and Ar. From these elements, only Cl and Ar
are not expected to be modified by the evolution of low and intermediate mass stars, and can be
used to infer the metallicity of the PN progenitor star. We could also use S but, as we mentioned
in Section~1, there are problems with the determination of the abundance of this element.
Figure~\ref{fig:ArCl} shows the Ar/Cl abundance ratio as a function of Cl/H for the group of PNe and
\ion{H}{ii} regions studied here. The small dots in the figure correspond to the values derived for
each Monte Carlo run and take into account all the uncertainties arising from errors in the line
fluxes involved in the calculations and those related to the adopted ICF. The metallicity of our
objects changes by an order of magnitude, but all of them show $\log(\mbox{Ar/Cl})\simeq1.1$ with
very low dispersion. The average value of the Ar/Cl abundance ratio considering our whole sample 
of ionized nebulae is $\log(\mbox{Ar/Cl})=1.10\pm0.02$, where 0.02 is the standard error of the mean. 
The tight correlation between the Cl/H and the Ar/H values is in agreement
with a lockstep evolution of both elements, at least for the metallicity range covered by our objects.
Since the Ar and Cl abundances are very difficult to calculate in the interstellar medium
\citep{jen09}, in stars, and in the Solar System \citep{asp09}, the results obtained from the analysis 
of photoionized nebulae provide the best estimates of the abundances of these two elements. 
 
The uncertainties in the values of Ar/H are significantly higher than those found for Cl/H, as can be
seen in Table~\ref{tab:5}. Hence, in what follows we use the Cl/H abundance ratio as a proxy for the
metallicity of all the objects.

\subsection{Metallicity differences between \ion{H}{ii} regions and PNe with CRD and ORD}

We also show in Figure~\ref{fig:ArCl} the protosolar abundances provided by \citet{lod10}.
Unfortunately, the Solar System abundances of these elements are not well constrained
\citep{asp09}. If we use the photospheric abundances provided by \citet{asp09} for Ar and Cl,
$12+\log(\mbox{Ar/H})=6.40$ and $12+\log(\mbox{Cl/H})=5.50$ with a correction of $\sim0.04$ dex due
to diffusion effects, we would get $12+\log(\mbox{Cl/H})=5.54$ and $\log(\mbox{Ar/Cl})=0.9$ instead
of the values favoured by \citet{lod10}, $12+\log(\mbox{Cl/H})=5.3$ and $\log(\mbox{Ar/Cl})=1.25$.
However, it is clear from Figure~\ref{fig:ArCl} that most of the ORD PNe have abundances close to or
above solar, \ion{H}{ii} region abundances are below solar, and CRD PNe show even lower abundances
(note that NGC~6826, our more metal-rich CRD PN, with $12+\log(\mbox{Ar/H})=6.28$ and
$12+\log(\mbox{Cl/H})=5.19$, is hidden behind the \ion{H}{ii} regions in this figure). It seems
that the progenitors of the CRD PNe in our sample were formed in a sub-solar metallicity medium, which
agrees both with the expectation of a higher efficiency of the TDU at low metallicities
\citep[see e.g.][]{kar14} and with the observation of larger fractions of CRD PNe in the metal-poor
environment of the Magellanic Clouds \citep{ber09}. On the other hand, the progenitors of the ORD PNe
were born in environments that cover our full metallicity range. In agreement with \citet{gar14},
we find that PNe with dual chemistry (with both silicates and PAHs in their spectra) have metallicities 
close to solar.

Several ORD PNe (7 out of 13) belong to the bulge, where the average PN metallicity is known to be
higher than in the disc \citep[see e.g.][]{wan07}. However, three of the disc PNe have higher values
of Cl/H than the \ion{H}{ii} regions. Since all our disc objects are located roughly at the same range
of Galactocentric distances, chemical evolution models predict that the PNe progenitor stars should
have formed at lower metallicities than the ones traced by \ion{H}{ii} regions. \citet{rod11} obtained
a similar result from an analysis of the oxygen abundances in a group of PNe and \ion{H}{ii} regions
belonging to the solar neighbourhood. \citet{rod11} speculated that this could be due to the presence
of organic refractory dust grains in \ion{H}{ii} regions containing significant amounts of oxygen, but
the fact that we find the same result for chlorine indicates that this is a more general effect. A
possible explanation is stellar migration of the central stars (and the Sun?) from inner regions of
the Galaxy, but other explanations, such as infall of metal-poor gas clouds into the Galaxy or changes
in the stellar composition arising during star formation or stellar evolution cannot be ruled out at
the moment. Six of the seven \ion{H}{ii} regions show similar metallicities, indicating that the
interstellar medium in the solar neighbourhood is chemically well mixed. This is confirmed by the
values we find for O/H for the seven \ion{H}{ii} regions (see Table~\ref{tab:5} and
Figure~\ref{fig:OCl} below). The chlorine abundance of NGC~3576 is somewhat low, but compatible within
the errors with the values of the other \ion{H}{ii} regions.

\subsection{Inconclusive evidence for neon enrichment}

\begin{figure}
\centering
\includegraphics[width=\hsize, trim = 30 15 40 0, clip =yes]{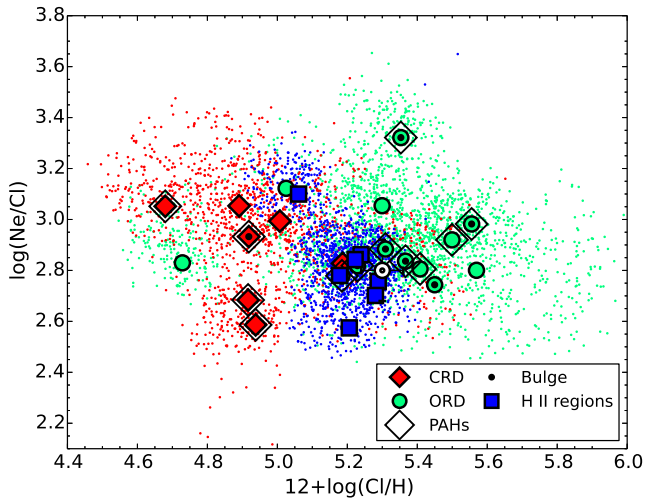}
\caption{Values of Ne/Cl as a function of Cl/H. The protosolar abundances of
\citet{lod10} are overplotted with the solar symbol. We use the same symbols as in Figure~\ref{fig:ArCl}.
\label{fig:NeCl}}
\end{figure}

Figure~\ref{fig:NeCl} shows the results for neon. Neon can be produced in stars with initial masses in
the range $\sim2$--4 \Msun\ \citep{kar03,pig13}, and \citet{mil10} argue that there is evidence for
neon enrichment in some Galactic PNe. Our Ne/Cl abundance ratios show a large dispersion, and the fact
that CRD PNe, \ion{H}{ii} regions, and ORD PNe cover similar ranges in Ne/Cl suggests that this
dispersion arises from uncertainties in the ICF for neon. Our sample PNe do not show clear evidence
for Ne enrichment, although we cannot rule it out since it could be masked by the uncertainties in the
derived Ne/H abundance ratio. If we reject the high value of Ne/Cl in NGC~3576, since it could be due
to the too low value found for Cl/H in this object, we can see that the other \ion{H}{ii} regions and
several PNe cluster around the solar value of Ne/Cl. The remaining PNe might have some neon
enrichment.

\subsection{Oxygen production in the PNe with CRD}

\begin{figure*}
\centering
\includegraphics[width=\hsize, trim = 20 15 40 0, clip =yes]{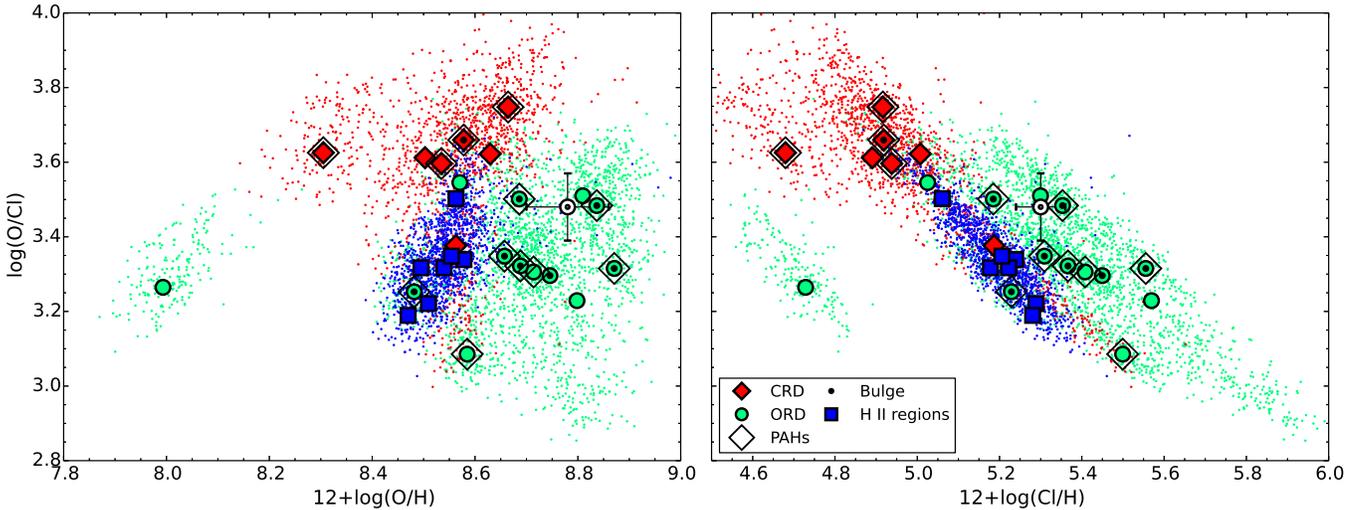}
\caption{Values of O/Cl as a function of O/H (left panel) and Cl/H (right panel) for our sample of ORD
PNe, CRD PNe, and \ion{H}{ii} regions. The protosolar abundances of
\citet{lod10} are overplotted with the solar symbol. We use the same symbols as in Figure~\ref{fig:ArCl}.
\label{fig:OCl}}
\end{figure*}

\begin{figure}
\centering
\includegraphics[width=\hsize, trim = 30 15 40 0, clip =yes]{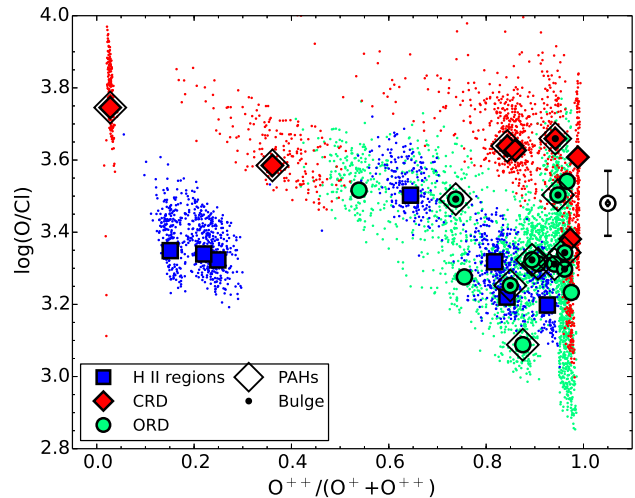}
\caption{Values of O/Cl as a function of O$^{++}$/(O$^{+}$+O$^{++}$). The protosolar abundances of
\citet{lod10} are overplotted with the solar symbol. 
We use the same symbols as in Figure~\ref{fig:ArCl}. \label{fig:OClgio}}
\end{figure}

Our results for the O/Cl abundance ratio as a function of O/H and Cl/H are presented in
Figure~\ref{fig:OCl}. Note that the trend followed by the dots that show the results of our Montecarlo
propagation of errors in the right panel of Figure~\ref{fig:OCl} is caused by the correlation between
the uncertainties in the values we derive for O/Cl and Cl/H. The trend is not present in our computed
values for these abundance ratios. However, in order to check if the ICF used to calculate the O/Cl
values is introducing a bias in our results, we plot in figure~\ref{fig:OClgio} the O/Cl abundance
ratios as a function of the degree of ionization, given by
$\omega=\mbox{O}^{++}/(\mbox{O}^{+}+\mbox{O}^{++}$), which is the parameter on which the ICF is based:
$\mbox{Cl/O}=(\mbox{Cl}^{++}/\mbox{O}^+)(4.1620-4.1622\omega^{0.21})^{0.75}$. We see that the adopted
ICF does not seem to be introducing any kind of bias in our results. In particular, the CRD PNe cover
the full range of values of $\omega$ and their O/Cl abundance ratios do not follow any trend with
$\omega$.

It can be seen in Figure~\ref{fig:OCl} that the metallicities implied by the Cl and O
abundances do not agree for all the objects. The Sun, the \ion{H}{ii} regions, and the ORD PNe show
similar values of the O/Cl abundance ratio, in agreement within the errors. However, the CRD PNe
have values of O/Cl which are clearly higher. The exception is NGC~6826, the CRD PN with the
highest metallicity, according to its chlorine abundance. The straightforward explanation of this
result is that the low-metallicity CRD PNe besides being C-rich are enriched in oxygen.

As we commented in Section~1, canonical AGB models like those of \citet{kar10} do not predict oxygen
enrichment at the metallicities that characterize our sample PNe, but models with diffusive convective
overshooting like those of \citet{pig13}, predict that the ejecta of stars with initial masses of
$\sim2$--3 \Msun\ and metallicities $Z=0.01$ and 0.02 will be enriched in both C and O (with C/O~$>1$),
with an increase in the O/H abundance ratio of up to $\sim0.26$ dex. This agrees with the differences
we find in the average values of $\log$(O/Cl) for our objects: $\log(\mbox{O/Cl})=3.65\pm0.06$ for the
CRD PNe excluding NGC~6826, $3.34\pm0.13$ for the ORD PNe, and $3.32\pm0.09$ for the 
\ion{H}{ii} regions. The fact that
the CRD PN with the highest metallicity, NGC~6826, has a value of O/Cl similar to the ones found for
ORD PNe and \ion{H}{ii} regions suggests that O enrichment is no longer important at metallicities
close to solar, but this should be confirmed by the analysis of other CRD PNe with similar
metallicity. Note that NGC~6826 is suspected to have a binary companion \citep{men89}, which might be
responsible for its carbon enrichment.

We performed a rough estimation of the amount of oxygen produced by low-mass stars and high-mass stars
using the yields calculated by \citet{pig13}. To do this, we integrated the net yields over the
\citet{sal55} initial mass function in two mass ranges: 1.65--5 \Msun\ and 15--60 \Msun\ for $Z=0.02$,
1.65--5 \Msun\  and 15--25 \Msun\ for $Z=0.01$. We obtain that low-mass stars contribute about 4 per 
cent and about 10 per cent of the total oxygen produced by a single stellar population at $Z=0.02$
and 0.01, respectively. These percentages are small, but they are likely to increase at lower
metallicities, and detailed chemical evolution models will be needed to find out if low-mass stars are
important producers of oxygen in different galaxies.

Figure~\ref{fig:OCl} offers an explanation for the overabundance of oxygen in PNe of the solar
neighbourhood found by \citet{rod11}. This result would be due to a combination of two causes: the
presence of high metallicity PNe in our vicinity for any of the reasons listed above plus oxygen
enrichment in some of the lower metallicity PNe.

\subsection{Nitrogen production in the PNe with CRD and ORD}

\begin{figure}
\centering
\includegraphics[width=\hsize, trim = 40 70 20 30, clip =yes]{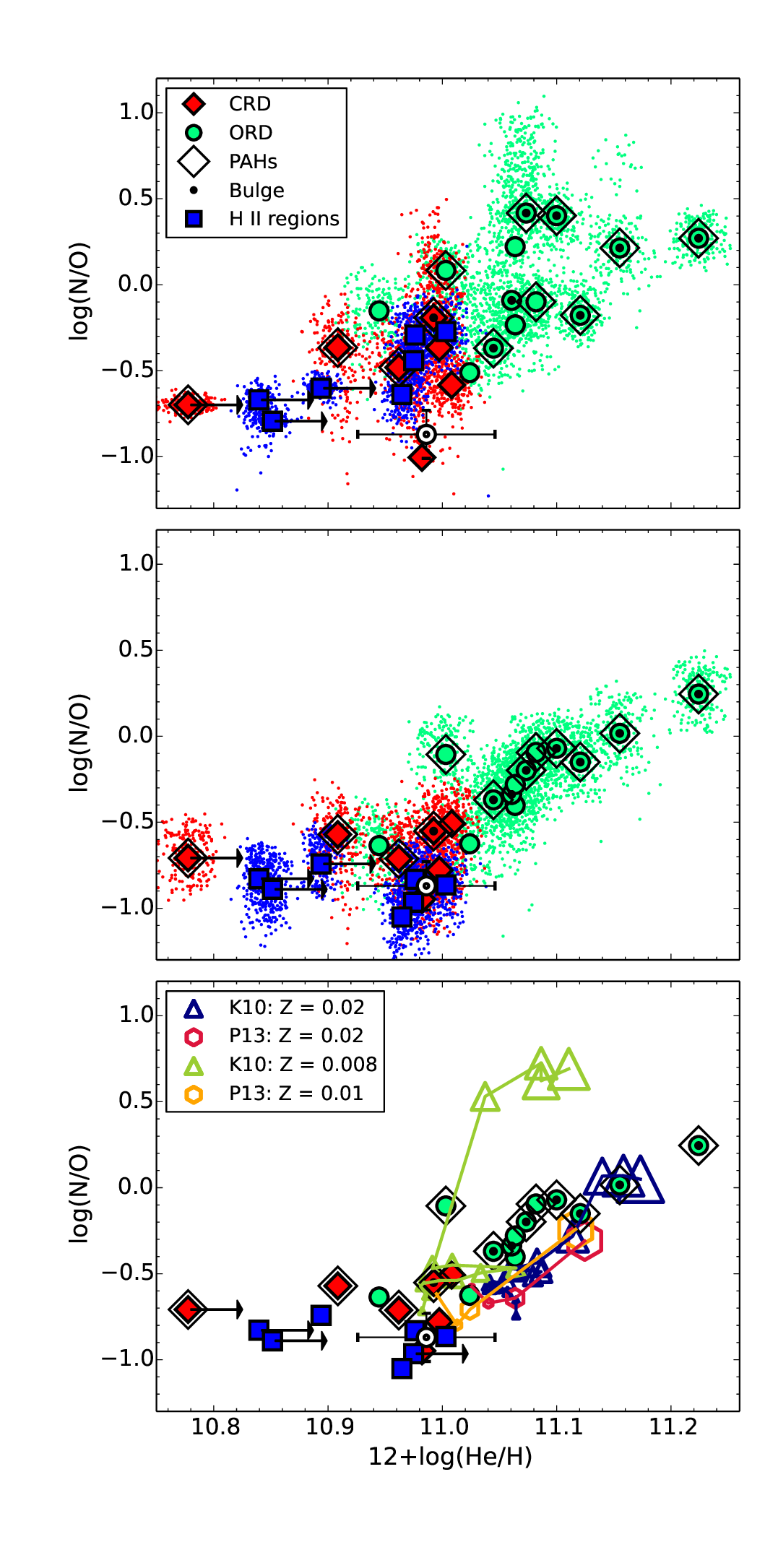}
\caption{Values of N/O as a function of He/H. The protosolar abundances of
\citet{lod10} are overplotted with the solar symbol. We use the same symbols as in Figure~\ref{fig:ArCl}. We
draw upper limits in the He abundance of four objects where the contribution of neutral helium is
significant. The top panel
shows the nitrogen abundances derived with the ICFs of \citet{del14b}; in the middle panel the
nitrogen abundances have been calculated using the classical ICF, N/O~$=\mbox{N}^+/\mbox{O}^+$. The
bottom panel shows the predictions from the nucleosynthesis models by \citet{kar10} and \citet{pig13}
overplotted for comparison with the results of the middle panel. The size of the symbols is larger for
higher mass progenitors (see the text for more information).\label{fig:NO}}
\end{figure}

Figure~\ref{fig:NO} shows the values of the N/O abundance ratio as a function of the helium abundance.
This plot is commonly used to identify the descendants of massive progenitors, which are efficient
producers of N and He according to the models of AGB evolution. The top panel of Figure~\ref{fig:NO}
shows the results obtained using the ICFs of \citet{del14b}. The \ion{H}{ii} regions show some
dispersion in their values of N/O and we noted that it seems to be related to their degree of
ionization, as measured with $\mbox{O}^{++}/(\mbox{O}^{+}+\mbox{O}^{++}$): the three \ion{H}{ii}
regions of higher ionization have the largest values of N/O. We decided to compare the results with those obtained using the classical
ICF for nitrogen: N/O~$=\mbox{N}^+/\mbox{O}^+$. The middle panel of Figure~\ref{fig:NO} shows the
abundances implied by this ICF. It can be seen that the new N/O ratios of the \ion{H}{ii} regions have
less dispersion and a better agreement with the solar value. On the other hand, the values of N/O in
the PNe with higher helium abundances now show a tight correlation with the helium abundance. We
cannot think of any observational effect or bias that would introduce this correlation if it were not
real, and we conclude that the classical ICF for nitrogen seems to be working better, at least for
this sample of objects. The issue merits more investigation, but since our conclusions do not depend
on the chosen ICF, we will not discuss this further.

The bottom panel of Figure~\ref{fig:NO} compares our results with the predictions of \citet{kar10} and
\citet{pig13}. The sequence of triangles with increasing size correspond to models with initial
metallicities of $Z=0.008$ and 0.02 and masses of 1, 1.25, 1.5, 1.75, 1.9, 2, 2.25, 2.5, 3, 3.5, 4,
4.5, 5, 5.5, 6, and 6.5 \Msun\ (the sequence ends at 6 \Msun\ for $Z=0.008$) computed by
\citet{kar10}. The sequence of hexagons correspond to the models of \citet{pig13} with initial masses
of 1.65, 2, 3, and 5 \Msun\ and $Z=0.01$ and 0.02. According to the models, most of our ORD PNe
descend from stars with masses above 4.5 \Msun, born in a medium of near-solar composition, whereas
the CRD PNe would descend from stars with masses below $\sim$4 \Msun\ formed in an environment with
sub-solar metallicity. Note however, that extra mixing processes, stellar rotation, or the effect of a
binary companion could produce large N/O ratios in PNe with initial masses not that high \citep{kar10,
sta13}. In particular, the models we are considering imply that all but one of the bulge PNe have
high-mass progenitors, as discussed by other authors \citep[see e.g.][]{wan07,gar14}. However, this
conclusion should be considered with caution. \citet{bue13} argues that if the bulge PNe had
progenitor masses above 4 \Msun, we would observe in the bulge bright carbon stars arising from
stars of 2--4 \Msun, which is not the case. Following the suggestion of \citet{nat12} that there is a
population of helium-enhanced stars in the bulge, similar to those found in some globular clusters,
\citet{bue13} constructs a set of AGB models with masses between 1.2 and 1.8 \Msun, $Z\sim0.02$, and
pre-enriched in helium, finding that they can reproduce the observed abundance ratios.

\begin{figure}
\centering
\includegraphics[width=\hsize, trim = 30 15 40 0, clip =yes]{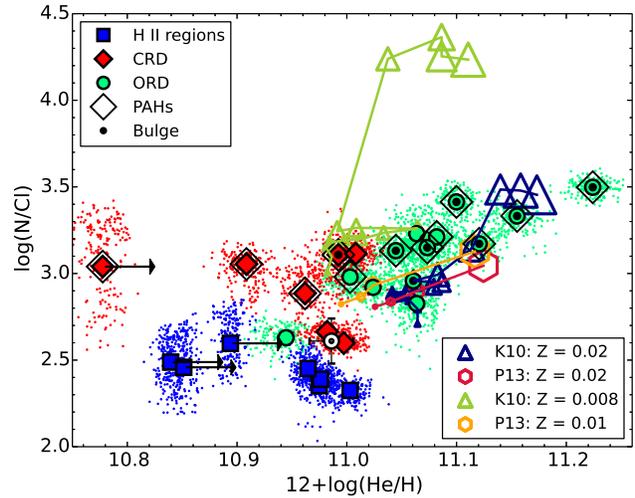}
\caption{Values of N/Cl as a function of He/H. The N abundances have been calculated using the
classical ICF, N/O~$=\mbox{N}^+/\mbox{O}^+$. The protosolar abundances of
\citet{lod10} are overplotted with the solar symbol. We use the same symbols as in Figure~\ref{fig:NO}. We
draw upper limits in the He abundance of four objects where the contribution of neutral helium is
significant. \label{fig:NCl}}
\end{figure}

Since chlorine is a better proxy for the metallicity than oxygen, we display in Figure~\ref{fig:NCl}
the results for the nitrogen abundances relative to chlorine.  We also show the predictions from 
the nucleosynthesis models of \citet{kar10} and \citet{pig13}. As expected, the results for the ORD
PNe are almost the same when compared to the values in \ion{H}{ii} regions because their oxygen
abundance has not been modified, but the N/Cl values obtained in the CRD PNe reveal that most 
of them are also significantly enriched in nitrogen. 

\subsection{The C/O abundance ratio}

\begin{figure*}
\centering
\includegraphics[width=\hsize, trim = 60 15 50 0, clip =yes]{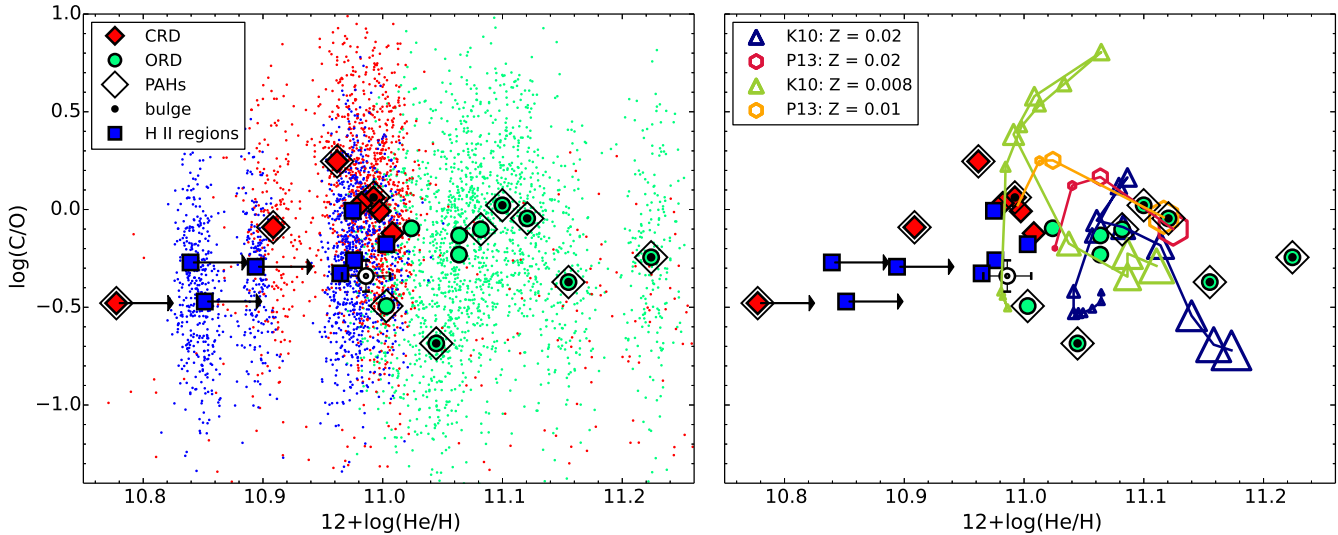}
\caption{Values of C/O as a function of He/H for the studied PNe and the \ion{H}{ii} regions. 
The protosolar abundances of \citet{lod10} are overplotted with the solar symbol. We use the same 
symbols as in Figure~\ref{fig:ArCl}. Note that for three of the ORD PNe we could not
estimate the value of C/O and these objects are not plotted in this figure. \label{fig:COHe}}
\end{figure*}

Figure~\ref{fig:COHe} shows the C/O values as a function of He/H, both derived from RLs. For three of
the ORD we do not have an estimate of the C/O abundance ratio and they do not appear in this figure. 
In principle, RLs and CELs might be tracing different nebular phases \citep[see e.g.][]{liu04b}, but the 
values of C/O derived using RLs and CELs are similar in most cases \citep{liu04b, del14a}.

We can see that the value derived for C/O does not provide a direct indication of the type of dust
grains present in a PN. This can be attributed to the uncertainties involved in its determination,
since the C/O value is based on weak RLs and requires an ICF. In fact, the dispersion of our Monte
Carlo error simulations shows that \ion{H}{ii} regions and ORD PNe have C/O values compatible with the
expectation that C/O~$<1$ in these objects and the values of CRD PNe are also compatible with
C/O~$>1$, as expected from their dust features. The average value of C/O for CRD PNe is also higher
than the one shown by the other objects, which cluster around the solar value. Another effect that
could affect the observed value of C/O is depletion of carbon or oxygen into dust grains. For these
reasons, the dust features observed in the infrared spectra are likely to provide more reliable
information on whether the value of C/O is below or above 1 in each object.

If we now examine the values of He/H (including those listed in Table~\ref{tab:5} for the three ORD
PNe that have no estimate of C/O), we can see that excluding the low-metallicity object from the halo,
the ORD PNe have helium abundances which are systematically higher than those found for CRD PNe and
\ion{H}{ii} regions. This is consistent with our conclusions above that they had massive (or
pre-enriched) stellar progenitors, since AGB models predict that helium enrichment takes place for the
stars with the highest masses. This is illustrated in the right panel of the figure, where we plot the
same models we used for Figure~\ref{fig:NO}.

\begin{figure}
\centering
\includegraphics[width=\hsize, trim = 20 15 50 0, clip =yes]{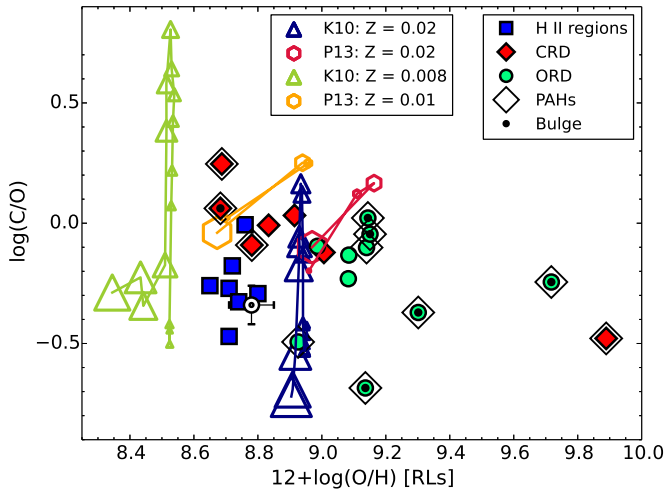}
\caption{Values of C/O as a function of O/H derived from RLs for the studied PNe and the \ion{H}{ii}
regions. The protosolar abundances of \citet{lod10} are overplotted with the solar symbol. We use the 
same symbols as in Figure~\ref{fig:NO}. Note that for three of the ORD PNe we
could not estimate the value of C/O and these objects are not plotted in this figure. \label{fig:COO}}
\end{figure}

In Figure~\ref{fig:COO}, we also show the derived values of C/O but now as a function of the oxygen
abundances implied by RLs (listed in Table~\ref{tab:5}). 
Since the correct values for the nebular oxygen abundances are likely to be intermediate between the
ones implied by CELs and RLs (or close to either one), the O/H values derived from RLs complete our
picture of the distribution in metallicity of the sample objects and their comparison with the Sun.
The O/H values for \ion{H}{ii} regions and ORD PNe should be corrected upward by 0.10--0.15 dex to
take into account the presence of oxygen in silicates and oxides within dust grains \citep{whi10, pei10}.
With this correction, the oxygen abundance in the \ion{H}{ii} regions would still be below solar for
the results based on CELs, but would agree both with the solar abundance and with the abundances found
for nearby B stars \citep{nie12} for the results based on RLs. On the other hand, the oxygen
abundances derived with CELs imply that the ORD PNe of the solar neighbourhood have a chemical
composition which is close to solar, whereas RLs imply that these objects are significantly more
metal-rich than the Sun. Both RLs and CELs imply that several nearby PNe are more metal-rich than the
\ion{H}{ii} regions, a result confirmed by the derived values of Cl/H. 

Figure~\ref{fig:COO} also shows the predictions of the AGB models of \citet{kar10} and \citet{pig13}.
In both sets of models the results for the lower mass objects indicate the initial values of O/H. As
we commented above, the models of \citet{kar10} for the most massive stars with $Z=0.008$ predict
significant oxygen destruction, whereas the models of \citet{pig13} with initial masses of 2--3 \Msun\
are enriched both in carbon and oxygen. As discussed above, our results for the CRD PNe agree with the
predictions of \citet{pig13}, at least for $Z=0.01$. An extended analysis of other CRD PNe might
provide more information.

\section{Conclusions}

We have studied the chemical composition of 20 Galactic PNe with high quality infrared and optical
spectra. The sample includes one halo PN, eight bulge PNe, and eleven disc PNe. We have also selected
from the literature seven \ion{H}{ii} regions from the solar neighbourhood  with the best available
optical spectra.Using the infrared dust features we have classified the PNe in two groups: PNe with
ORD and PNe with CRD. The first ones show amorphous and/or crystalline silicates in their spectra
whereas the second ones show the broad features at 11 and/or 30 $\mu$m associated with SiC and MgS.

We have computed the physical conditions and the chemical abundances following the same procedure in
all the objects. The abundances derived here are based on the best available ICFs, mainly those
derived by \citet{del14b}, and the uncertainties in the element abundances include the uncertainties
introduced by the physical conditions, the ionic abundances, and the ICFs. 

We find a tight correlation between the abundances of Cl and Ar, in agreement with a lockstep
evolution  of both elements. The metallicity in our sample, traced by Cl/H, covers one order of magnitude, 
and there are significant differences in
the Cl/H ratio among the three groups of objects studied here. In general, the PNe with CRD have low
metallicity, the \ion{H}{ii} regions have intermediate metallicities, and the PNe with ORD cover the
whole range of metallicity (with some of them showing the highest abundances, even higher than the
solar ones). This result indicates that the progenitors of the PNe with CRD were formed in a sub-solar
metallicity medium (with the exception of NGC~6826, with a value of Cl/H similar to the \ion{H}{ii}
regions) whereas most of the PNe with ORD were formed in a  near-solar metallicity medium (with the
exception of the halo PN DdDm~1). 

The neon abundances at a given metallicity show a relatively large dispersion. The dispersion could 
be due to uncertainties in the ICF but we cannot rule out that some of the PNe are neon enriched. 

The oxygen abundances do not always trace the metallicities of the objects. The O/Cl values in the
Sun, the PNe with ORD, and the \ion{H}{ii} regions are similar, but most of the PNe with CRD show O/Cl
values that are higher by a factor of two. The exception is NGC~6826, the CRD PN with higher
metallicity in our sample. As we mentioned above, the progenitors of most of the PNe with ORD were
formed in an environment with near-solar metallicity, where the nucleosynthesis process that can
change the initial oxygen abundance are less efficient. This is in agreement with the similar values
of O/Cl found in these PNe and in the \ion{H}{ii} regions. The low Cl/H abundances and the high O/Cl
values found in all but one of the PNe with CRD provide evidence of oxygen production in the
progenitor stars of PNe with CRD. Non-standard nucleosynthesis models, such as those by \citet{pig13},
that include an extra mixing mechanism, predict a significant production of oxygen via the third
dredge-up in stars with masses around $\sim2$--3 \Msun, explaining our results. The relevance for
galactic chemical evolution of oxygen production in low mass stars deserves further study using
chemical evolution models. 

The comparison of the He/H, C/O, and N/O abundance ratios derived for the PNe with
the predictions of nucleosynthesis models, suggests that the PNe with CRD descend from stars with
masses in the range 1.5--3 \Msun whereas the PNe with ORD and the highest abundances of nitrogen and
helium would arise from stars with higher initial masses. The theoretical models indicate that they
descend from stars with masses $\gtrsim4.5$ \Msun, but this limit would be smaller with models that
include stellar rotation, extra mixing processes or pre-enrichment.

Our results indicate that Ar and Cl are the best metallicity indicators for PNe. At any metallicity 
their abundances reflect the composition of the interstellar medium where the PN progenitors 
were formed. Moreover, the ICFs adopted here for Ar and Cl can be used when only Cl$^{++}$ 
and Ar$^{++}$ lines are observed, making easier a homogeneous calculation of PN metallicities. 
High resolution and deep spectra are needed, especially of extragalactic PNe, to delve in this 
issue and calculate, for example, metallicity gradients using Cl or Ar. Besides, the obtention of 
high quality optical spectra in which we can measure Cl and Ar lines will provide more reliable 
estimates of the electron densities. The observed extragalactic PNe are often the brightest ones, 
with high electron densities, where the commonly used density diagnostic ratio based on [\ion{S}{ii}] lines 
is not sensitive.

\section*{Acknowledgements}
We are grateful to the anonymous referee for valuable comments and suggestions.
The authors thank L.~Carigi, L.~Hern\'andez-Mart\'inez, and M.~Pignatari for fruitful discussions 
and comments. G.D.-I. gratefully acknowledges a DGAPA postdoctoral grant from the 
Universidad Nacional Aut\'onoma de M\'exico (UNAM). C.M. and G.S. acknowledge support 
from the following Mexican projects: CB-2010/153985, PAPIIT-IN105511, and PAPIIT-IN112911.
M.P. is grateful for the finantial support provided by CONACyT grant 129753. 
M.R. and G.D.-I. acknowledge support from Mexican CONACyT grant CB-2009-01/131610. 
This work has 
made use of NASA's Astrophysics Data System, and the SIMBAD database operated at CDS, 
Strasbourg, France. 

{}
\label{lastpage}

\end{document}